\documentclass[11pt]{article}
\usepackage{epsfig}
\usepackage{amssymb,amsmath}
\usepackage{epigraph}
\usepackage{xcolor}

\usepackage[lined,boxed,commentsnumbered]{algorithm2e}

\usepackage{graphicx}				
\usepackage{amssymb}
\usepackage{multirow}

\usepackage{siunitx,booktabs}
\usepackage{newtxtext,newtxmath}

\DeclareMathOperator\erfc{erfc}
\DeclareMathOperator\erfcx{erfcx}

\newcommand{\beq}{\begin{equation}}
\newcommand{\eeq}{\end{equation}}
\newcommand{\bea}{\begin{eqnarray}}
\newcommand{\eea}{\end{eqnarray}}

\begin{document}



\begin{center}

{\LARGE
Non-Equilibrium Skewness, Market Crises, and Option Pricing: 
\vskip0.5cm
Non-Linear Langevin Model of Markets with Supersymmetry   

}

\vskip1.0cm
{\Large Igor Halperin\footnote{Fidelity Investments. E-mail: igor.halperin@fmr.com. Opinions expressed here are author's own, and do not represent views of his employer. A standard disclaimer applies. E-mail for communications on the paper: ighalp@gmail.com}  
} \\
\vskip0.5cm
First version: November 8, 2020 \\
This version: \today \\

\vskip1.0cm
{\Large Abstract:\\}
\end{center}
\parbox[t]{\textwidth}{
This paper presents a tractable model of non-linear dynamics of market returns using a Langevin approach.
Due to non-linearity of an interaction potential, the model admits regimes of both small and large return fluctuations.
Langevin dynamics are mapped onto an equivalent quantum mechanical (QM) system. 
Borrowing ideas from supersymmetric quantum mechanics (SUSY QM), a parameterized  ground state wave function (WF) of this QM system 
is used as a direct input to the model, which also fixes a non-linear Langevin potential. 
Using a two-component Gaussian mixture as a ground state WF with an asymmetric double well potential produces a tractable low-parametric model 
with interpretable parameters, referred to as the NES (Non-Equilibrium Skew) model.  
Supersymmetry (SUSY) is then used to find time-dependent solutions of the model in an analytically tractable way.
Additional approximations give rise to a final practical version of the NES model, where real-measure and risk-neutral return distributions are given by 
three component Gaussian mixtures. This produces a closed-form approximation for option pricing in the NES model by a mixture of three Black-Scholes prices,
providing accurate calibration to option prices for either benign or distressed market environments, while using only a single volatility parameter.
These results stand in stark contrast to the most of other option pricing models such as local, stochastic, or rough volatility models that need more complex specifications of noise to fit the market data.   

 }
 
 \newcounter{helpfootnote}
\setcounter{helpfootnote}{\thefootnote} 
\renewcommand{\thefootnote}{\fnsymbol{footnote}}
\setcounter{footnote}{0}
\footnotetext{
I would like to thank Damiano Brigo, John Dance, Cris Doloc, Lisa Huang and Yinsen Miao for comments and helpful remarks. 
}     

 \renewcommand{\thefootnote}{\arabic{footnote}}
\setcounter{footnote}{\thehelpfootnote} 

\newpage
 
\section{Introduction}

This paper presents an analytically tractable model of non-linear dynamics of market returns using a Langevin approach, with five free parameters. 
Market returns are driven by a combination of a non-linear \emph{drift potential} (a function of returns) controlled by four parameter $ \mu, \sigma_1, \sigma_2, a $, and a Gaussian white noise, whose strength is given by a volatility parameter $ h $.
Due to a non-linear potential, the model admits regimes of both small and large return fluctuations. 
A stationary distribution of the model is given by a function that only involves the first four parameters $ \mu, \sigma_1, \sigma_2, a $, while the volatility 
parameter $ h $ controls pre-equilibrium, time-dependent effects. Both the stationary distribution and a time-dependent transition probability density are analytically tractable and available in closed form or in quadratures. The model produces closed-form approximations for option pricing and market crash or crises probabilities.
In contrast to a vast majority of the option pricing literature that resort to complex models of the noise such as local, stochastic, or rough volatility models, 
the model presented in this paper achieves an accurate fit to market option prices using only a single volatility parameter $ h $. 
\emph{Mutatis mudandis}, the model can also be applied to individual stocks.
 
The return distribution obtained with the model can be represented as a sum of a stationary distribution and a time-varying component. The stationary distribution is 
a \emph{direct input} to the model, in contrast to more traditional approaches where a stationary distribution is normally an output, rather than an input to a model. In this paper, the stationary distribution is given by the square of a two-component Gaussian mixture (which produces a \emph{three}-component Gaussian mixture), defined using four parameters  
$ \mu, \sigma_1, \sigma_2, a $. 

While a stationary return distribution is an \emph{input} to our model, the \emph{output} is given by a time-varying, dynamic component of the return distribution obtained with the model.
The dynamic component is driven by \emph{both} the first four parameters  $ \mu, \sigma_1, \sigma_2, a $ and the fifth 
parameter $ h $. The latter controls probabilities of large market drops or crises within a certain time horizon, e.g. a year, and hence drives the market skewness (and other moments) in a non-equilibrium setting. We will therefore refer to the model presented below as the Non-Equilibrium Skew (NES) model.

A decomposition of the return distribution into a stationary and a time-varying component is very convenient because it allows one, among other things, to also decompose moments of the distribution such as variance or skewness into a constant and time-varying components. While substantial time variations of moments of return distributions in financial markets are well documented phenomena, models that try to fit or explain such variations usually do not introduce such a structural separation. Instead, they follow purely statistical or CAPM-based regression-based approaches (see e.g. \cite{Harvey_2000}), or use option pricing models to try to infer market views on future variance or skewness from option quotes, see e.g. \cite{Stilger}. Option-based approaches usually infer moments of a risk-neutral (option-price implied) return distribution rather than a real return distribution, and additional modeling steps are required to relate the two distributions. 

 The model presented in this paper constructs a direct link between 
deep out-of-the-money (OTM) options and moments of future returns under \emph{both} risk-neutral and real-measure probability measures. 
By calibration to option data, the model can identify both a constant and time-varying contributions to the skew and variance of market returns. 
The empirical finance literature established contemporaneous correlations between a state of economy (e.g. proxied by the S\&P 500 returns) and the returns' skewness, as well as explored ways to use the historical or option-implied market skewness or variance as predictors of future returns, see e.g. \cite{Ferreira_2018,Stilger}. This paper offers a theoretical framework to pursue both sorts of analysis, with a clear separation between a static and dynamic components of the problem. 

The NES model can be easily calibrated to market data for European vanilla options using a simple analytical approximation.
Remarkably, while the development of the model involves some advanced concepts from quantum mechanics as will be explained below, its final version recommended for a practical use is very simple, and is given by two three-component Gaussian mixtures as models for the risk-neutral and real-measure return distributions, where parameters for both mixtures are fixed in terms of the original model parameters $ \mu, \sigma_1, \sigma_2, a, h $. 
Respectively, the NES model  produces options prices as \emph{mixtures of three Black-Scholes prices}.     
 Numerical experiments show that the model is flexible enough to both calibrate to option data and match typical levels of market volatility and skewness across different market regimes.

The model developed here is based on a Langevin approach where we directly model non-linear stochastic dynamics of market log-returns, denoted $ y_t $ in this paper. With the Langevin approach, dynamics are defined in terms of an interaction potential function $ V(y_t) $ and a noise term.
For the latter, we use a simplest assumption of a constant Gaussian noise with volatility $ h $, though other more complex specifications (e.g. a state-dependent Gaussian noise, a non-Gaussian noise, or various multivariate extensions) could also be considered. For the former, we choose a potential function $ V(y) $ for which a stationary distribution $ p_s(y_t) $ of log-returns has a known and \emph{fixed} functional form, and is given by a three-component Gaussian mixture mentioned above. In other words, a potential $ V(y) $ in this paper is constructed in such a way that the problem becomes \emph{quasi-exactly solvable}. Here the word 'quasi' refers to the fact that while the stationary distribution $ p_s(y_t) $ is an analytical expression, and is actually an input to our model, the model output amounts to computing the \emph{dynamic} behavior - and this requires some computational efforts, and relies on certain approximations. 

To develop an analytically tractable formulation, with easily computable approximations that could also be improved to any required precision, we borrow ideas from physics. First, we use a hidden supersymmetry of Langevin dynamics
that was discovered in the 1980s \cite{Feigelman_Tsvelik}.  The Langevin dynamics are mapped onto an equivalent quantum mechanical (QM) system, where 
the role of the Planck constant $ \hbar $ is played by the square of the volatility parameter $ h^2 $ (which explains our choice for this variable). In such an equivalent QM system, the hidden supersymmetry (SUSY) of the original Langevin dynamics becomes manifest, see e.g. \cite{Junker}. The resulting equivalent QM system is mathematically identical to an Euclidean version of supersymmetric quantum mechanics (SUSY QM) developed by Witten \cite{Witten_SUSY} in 1981. While SUSY QM was originally proposed as a toy model to study supersymmetry in quantum field theory and string theory, since then it has become a topic of research on its own. In particular, researchers in SUSY QM 
found that SUSY can be used to develop very efficient computational schemes for computing wave functions or energies of different QM systems, that are often both simpler and more efficient than more traditional methods of computing in quantum mechanics, see e.g. \cite{SUSY_QM_Cooper} or \cite{Junker}. This paper proposes to use this machinery to model non-linear stochastic dynamics of market returns. 
 
 Further using ideas from SUSY QM  \cite{SUSY_QM_Cooper}, we use a parameterized  ground state wave function (WF) of this QM system as a \emph{direct input} to the model, which also fixes a non-linear Langevin potential $ V(y) $. 
A stationary distribution of the original Langevin model is given by the square of this WF, and thus is also a direct input to the model.  
This ensures that the model is quasi-exactly solvable. We 
use a two-component Gaussian mixture with parameters $ \mu, \sigma_1, \sigma_2, a $ as a model of a ground state WF with an asymmetric double well potential, one of the most canonical examples in physics \cite{Landau_QM, Zinn-Justin-QFT}.  This parametrization produces a tractable low-parametric model  with intuitive and interpretable parameters,  
where supersymmetry (SUSY) is used to find time-dependent solutions of the model in an analytically tractable way.   

Note that while a Langevin model with a non-linear potential
could be introduced based on phenomenological grounds, the approach of this paper is motivated by previous work in \cite{HD_QED} and \cite{Inverted_World}
that developed a non-linear Langevin model of stock price dynamics.\footnote{Langevin dynamics with a cubic potential was previously considered by Bouchaud and Cont \cite{BC}, however for a different dependent variable (the speed of a market price),
and applied to modeling of market bubbles and crashes. Related ideas were also discussed by Sornette \cite{Sornette_book}.} 
A model developed in \cite{HD_QED,Inverted_World}, called the QED (``Quantum Equilibrium-Disequilibrium") model, explains non-linearities of price dynamics as a combined effect of capital inflows or outflows in the market, and their impact on asset returns, modeled as a linear or quadratic function of capital inflows. 
While the approach of \cite{HD_QED, Inverted_World} provides a theoretical motivation for considering non-linear Langevin dynamics, here 
we present a different formulation that operates with similar non-linear potentials to those presented in \cite{HD_QED, Inverted_World}\footnote{in the sense that both specifications may produce either a single well or a double well potential, depending on the parameters.}, but is more tractable analytically. More specifically, the potential used in this paper is 'semi-phenomenological', and is inspired by a double-well potential, one of the most famous potentials in quantum and statistical mechanics and quantum field theory, that serves there as a prime model for describing quantum tunneling and thermally induced barrier transitions \cite{Landau_QM, Zinn-Justin-QFT}.\footnote{The approach of this paper can also be viewed as partially inspired by machine learning paradigms, because it relies on a Gaussian mixture parametrization of a prime modeling objective - which in our case is given by a ground state wave function $\Psi_0 $. A similar approach to the one developed in this paper can also be applied to other dynamical systems.}

Theoretical links notwithstanding, non-linear Langevin models for market returns can also be considered on purely phenomenological  grounds.  Note that squares of market returns are sometimes used in the literature as additional cross-sectional predictors of stock returns in quadratic extensions of the CAPM model \cite{CAPM}, also referred to as 'quadratic CAPM' models,  see e.g. \cite{Chabi-Yo-2007} and references therein. 

In this paper, squares and higher powers of market returns are used for time-series modeling.
The main novelty of our Non-Equilibrium Skew (NES) model is that non-linearities of return \emph{dynamics} give rise to certain solutions that are simply impossible to obtain
starting with a linear model, or even with a non-linear model but treated using methods of perturbation theory. Other, \emph{non-perturbative} (i.e. not traceable within a perturbation theory) saddle-point solutions of the dynamics called \emph{instantons} become important to capture the right dynamics of the model.
While instantons are encountered in many problems in statistical physics and quantum field theory (see e.g. \cite{Zinn-Justin-QFT}), the QED model of \cite{HD_QED,Inverted_World}
 suggest that instantons might also be relevant for modeling of market prices and returns. This paper offers further insights into the importance of instantons (and,  in general, of a non-linear and non-perturbative analysis) for modeling dynamics producing both small and large market fluctuations - with the latter interpreted as events of a severe market downturn or a crisis. More specifically, instantons and other non-linear effects induce both market crises and market skewness, enabling a theoretical link between them.
   
The paper is organized as follows. Sect.~\ref{sect_factor_dynamics} presents the Langevin model along with its Fokker-Plank and  Schr{\"o}dinger 
representations that produce an equivalent quantum mechanical (QM) system. Sect.~\ref{SUSY} introduces supersymmetry (SUSY) as a tool to solve the QM system, and 
Sect.~\ref{sect_Gaussian_mixture_WF} presents a quasi-exactly solvable NES model for a Langevin potential motivated by a double well potential in QM.
Instantons and their role in dynamics are discussed in Sect.~\ref{Instantons_and_Kramers}. Sect.~\ref{SUSY_higher_states} applies SUSY to compute pre-asymptotic corrections to a stationary distribution of the model.\footnote{This section can be skipped at the first reading, as the next Sect.~\ref{sect_Applications}
can be read largely independently of Sect.~\ref{SUSY_higher_states}.}
 Sect.~\ref{sect_Applications} presents a simplified practical implementation of the NES model, and then applies it to explore time variations in return moments, and to option pricing out of equilibrium. Sects.~\ref{sect_option_pricing} and \ref{sect_option_pricing_2} show that pricing of European options in the NES model reduces to a three-component mixture of Black-Scholes option prices, where all parameters are fixed in terms of the original model parameters.  
 The following Sect.~\ref{sect_calib_to_SPX_options} presents examples of calibration to  SPX options.
The final Section~\ref{sect_Summary} concludes. Additional materials are presented in Appendices A to D.

\section{Langevin dynamics, double well potential, SUSY and the NES}
\label{sect_factor_dynamics}

\subsection{Langevin dynamics: the Fokker-Planck  and Schr{\"o}dinger pictures}

Let $ S_t $ be the value of a market index at time $ t $. A period-$T$ log-return  $ y_t $ is defined as follows 
\beq
\label{y_t_S_t_0}
y_t = \log \frac{S_t}{S_{t-T}}
\eeq 
In this paper, we focus on modeling dynamic distributions of $ y_t $ rather than distributions of market value (or a stock price) $ S_t $. The reason
is that unlike price dynamics which are non-stationary due to a market or stock drift, returns dynamics can be either stationary or non-stationary. If we define dynamics in terms of 
log-returns $ y_t $ instead of prices $ S_t $, we can differentiate between regimes of stationary vs non-stationary returns, which would be impossible if a model is formulated in terms of prices $ S_t $. As an interplay between these regimes is a key ingredient of analysis in this paper, we explore dynamics of log-returns 
$ y_t $ rather than prices $ S_t $. On the other hand, as $ y_T = \log(S_T/S_0) $, knowing a time-$T$ distribution of $ y_T $ also gives a distribution of $ S_T $.
 
We use the Langevin approach to describe continuous-time stochastic dynamics of market returns $ y_{t} $, where dynamics are determined by   
an (overdamped) Langevin 
 equation \cite{Langevin}:
 \beq
 \label{Langevin_return_1M}
 d y_{t} = - \frac{ \partial V (y_{t})}{\partial y_{t} } dt + h d W_t
 \eeq
where $ h $ is a volatility parameter, $ W_t $ is a standard Brownian motion, and $ V(y) $ is an interaction potential. In general, $ V(y) $ is a non-linear function whose structure is either deduced from underlying microscopic dynamics, or alternatively postulated based on phenomenological grounds. In particular, a quadratic potential 
$ V(y) = \frac{1}{2} \omega^2 (y-y_{\star})^2 $ gives rise to a linear drift $ \mu(y) = - \partial V/ \partial y $, producing a familiar Ornstein-Uhlenbeck (OU) process
as a special choice for the Langevin dynamics (\ref{Langevin_return_1M}). 
While a different specification of the interaction potential $ V(y) $ with higher-order non-linearities  will be presented below, for now we will proceed with a general development applicable for an arbitrary potential $ V(y) $. The only assumptions used in this section is that a minimum value of the potential is zero, i.e. 
$ V(y_{\star}) = 0 $ where $ y_{\star} $ stands for a minimum point of $ V(y) $, and that the potential $ V(y) $ rises as $ y \rightarrow \pm \infty $ at least as fast as 
$ V(y) \sim y^2 $. The last assumption is needed to ensure the existence of a stationary solution to the dynamics.

Transition probabilities $ p(y,t | y_0) $ to be at a certain state $ y = y_t $ at time $ t $ starting with an initial position $ y_0 $ at time $ t = 0 $ satisfy a Fokker-Planck equation (FPE) corresponding to
 the Langevin equation (\ref{Langevin_return_1M})
 \beq
\label{FPE}
\frac{\partial p(y,t | y_0)}{\partial t} =  \frac{\partial}{\partial y} \left[ \frac{\partial V(y)}{\partial y} p(y,t | y_0) \right] + \frac{h^2}{2} \frac{\partial^2}{\partial y^2} 
p(y,t | y_0)
\eeq
Alongside the transition probability density $ p(y,t | y_0) $ that satisfies the FPE (\ref{FPE}), it proves useful to consider a related transition density obtained if we used an inverted potential $ - V(y) $ in the Langevin equation (\ref{Langevin_return_1M}). Denoting such transition density $  p_{+} (y,t | y_0) $ and using the notation  $  p_{-} (y,t | y_0) $ for the solution of the original FPE 
(\ref{FPE}), the pair $ p_{\pm} (y,t | y_0) $ satisfies a pair of Fokker-Plank equations that can be compactly written as follows:
\beq
\label{FPE_tilde}
\frac{\partial p_{\pm}(y,t | y_0)}{\partial t} = \mp \frac{\partial}{\partial y} \left[  \frac{\partial V(y)}{\partial y} p_{\pm}(y,t | y_0) \right] + \frac{h^2}{2} \frac{\partial^2}{\partial y^2} 
p_{\pm}(y,t | y_0)
\eeq
with the initial condition $ p_{\pm}(y,0 | y_0) = \delta( y - y_0) $. We use the following ansatz to solve Eqs.(\ref{FPE_tilde})  (see e.g. \cite{vanKampen}):
\beq
\label{ansatz}
 p_{\pm} (y,t | y_0) =  e^{ \pm \frac{V(y) - V(y_0) }{h^2} }  \Psi_{\pm}(y,t | y_0).
\eeq
Using this in Eq.(\ref{FPE_tilde}) produces two imaginary time Schr{\"o}dinger equations (SE) for 
$  \Psi_{\pm}(y,t | y_0) $: 
\beq
\label{SE}
- h^2 \frac{\partial \Psi_{\pm}(y,t | y_0)}{\partial t} = \mathcal{H}_{\pm} \Psi_{\pm} (y,t | y_0),
\eeq
where $ h^2 $ serves as a Planck constant $ \hbar $, and $ \mathcal{H}_{\pm} $ are the Hamiltonians 
\beq
\label{Hamiltonian}
\mathcal{H}_{\pm} = -  \frac{h^4}{2} \frac{\partial^2}{\partial y^2} +  \frac{1}{2} \left( \frac{ \partial V}{\partial y} \right)^2 
\pm \frac{1}{2}  h^2 \frac{\partial^2 V}{\partial y^2}  \equiv -  \frac{h^4}{2} \frac{\partial^2}{\partial y^2} + U_{\pm}(y). 
\eeq
Note that the Hamiltonian $ \mathcal{H}_{-} $ transforms into 
the partner Hamiltonian $ \mathcal{H}_{+} $ if we flip the sign of 
the potential $ V(y) \rightarrow - V(y) $. Further properties of the pair of Hamiltonians $ \mathcal{H}_{\pm} $ will be explored in the next section, while here 
we focus on the 'prime' Hamiltonian $ \mathcal{H}_{-} $ corresponding to the initial FPE (\ref{FPE}).  

Let $ \left\{ \Psi_n^{-} \right\} $ be a complete set of eigenstates of the Hamiltonian $ H_{-} $ with eigenvalues 
$ E_n^{-} $:
\beq
\label{Eigen_values} 
\mathcal{H}_{-} \Psi_n^{-} = E_n^{-} \Psi_n^{-}, \; \; \; 
\sum_{n} \Psi_n^{-}(x) \Psi_n^{-}(y) = \delta(x-y).
\eeq
We assume here a discrete spectrum corresponding to a 
motion in a bounded domain, so that the set of values 
$ E_{n}^{(-)} $ is enumerable by integer values $ n = 0, 1, \ldots $. We can now expand the wave function of the time dependent SE (\ref{SE}) in stationary states 
(quantum-mechanical wave functions, or WF's)
 $ \left\{ \Psi_n^{-} \right\} $:
 \beq
 \label{expansion_in_stationary_states}
 \Psi_{\pm}(y,t | y_0)  = \sum_{n=0}^{\infty} c_n^{\pm}  e^{ - \frac{ t E_n^{-}}{h^2}}    \Psi_n^{-}(y) 
 \eeq
 where coefficients $ c_n^{\pm} $ should be found from the initial condition on $ \Psi_{\pm}(y,t | y_0) $ which can be read off the initial condition 
  $ p_{\pm}(y,0 | y_0) = \delta( y - y_0) $ and Eq.(\ref{ansatz}):
  \beq
  \label{c_n} 
  c_n^{\pm} = \int_{-\infty}^{\infty}  \Psi_n^{-}(y)  \Psi_{\pm}(y,0 | y_0) =  e^{ \pm \frac{V(y) - V(y_0) }{h^2}} 
  \eeq
 Combining Eqs.(\ref{ansatz}), (\ref{expansion_in_stationary_states}) and (\ref{c_n}), we obtain the spectral decomposition of the original FPE (see e.g. \cite{vanKampen}):
 \beq
\label{spectral_FPE}
 p(y,t | y_0) =  e^{ - \frac{V(y) - V(y_0)}{h^2}}  \sum_{n=0}^{\infty} e^{ - \frac{ t E_n^{-}}{h^2}} 
 \Psi_n^{-}(y) \Psi_n^{-}(y_0).
 \eeq
When $ t \rightarrow \infty $, only one term with the lowest energy 
survives in the sum in Eq.(\ref{spectral_FPE}). As the potential $ U_{-}(y) $ is bounded below by zero, the lowest state, if it exists, would be a state 
$  \Psi_0^{-}(y) $ with zero energy $ E_0^{-} = 0 $. In our setting, we assume that a potential $ V(y) $ is such that a normalizable zero energy state 
$  \Psi_0^{-}(y) $ does exist. Retaining only two leading terms in the expansion (\ref{spectral_FPE}) and defining $ \Delta E := E_1^{-} - E_0^{-}  =  E_1^{-} $ as an energy splitting between the lowest and the first excited states, we can write it in the following form:
\beq
\label{two_terms_approx}
p(y,t | y_0) =  e^{ - \frac{V(y) - V(y_0)}{h^2} }  \left[  \Psi_0^{-}(y) \Psi_0^{-}(y_0) + 
e^{- \frac{ t  \Delta E}{\sigma^2}}   \Psi_1^{-}(y) \Psi_1^{-}(y_0) + \ldots \right] 
\eeq
The last expression produces both time-dependent and stationary probability densities for the original problem corresponding to the Langevin equation 
(\ref{Langevin_return_1M}) and the FPE (\ref{FPE}).\footnote{If a lowest energy state has a positive energy   
 $ E_0^{(-)}  > 0 $, this would modify Eq.(\ref{two_terms_approx}) by a time-decay factor $ e^{ - t   E_0^{(-)}/h^2} $, indicating a   
a metastability of an initial state $ y_0 $ that would decay with the decay rate 
$ E_0^{(-)}/h^2 $.} 
 In particular, a stationary distribution is obtained by taking a limit $ t \rightarrow \infty $, where only the first term survives.\footnote{An apparent remaining dependence on the initial position $ y_0 $ in the limit $ t \rightarrow \infty $ of Eq.(\ref{two_terms_approx}) is spurious and will be shown to cancel out in the next section.} On the other hand, for finite values $ t < \infty $, Eq.(\ref{two_terms_approx}) describes a pre-asymptotic behavior 
en route to a stationary state, where a time-dependent correction is given by the second term in (\ref{two_terms_approx}). At yet shorter times, higher-order corrections omitted in (\ref{two_terms_approx}) may become important.

The spectral representation (\ref{spectral_FPE}) (or its approximation (\ref{two_terms_approx})) that gives a decomposition of the solution of the FPE 
(\ref{FPE}) in terms of eigenvalues of the SE (\ref{SE}) is well known in the literature. With a conventional approach, one starts with defining a potential 
$ V(y) $, which is then used to find a Schr{\"o}dinger Hamiltonian $ \mathcal{H}_{-} $, and then the latter is used to compute eigenvalues $ \Psi_n^{-} $.
In the next section, we will present an alternative approach that provides a more direct link between the FPE transition density $ p(y,t | y_0) $ and 
wave functions $ \Psi_n^{-} $. As will be shown next, the quantum-mechanical wave functions  $  \Psi_n^{-} $ serve as 'half-probabilities', in the sense that 
the FP density  $ p(y,t | y_0) $ is related to squares of $ \Psi_n^{-} $, in a close analogy to how a square of a wave function in the conventional quantum mechanics is interpreted as a probability density of a quantum mechanical particle.
   
\subsection{Dynamics from statics: SUSY and half-probability distributions}
\label{SUSY}

To obtain a different representation of an approximate spectral decomposition (\ref{two_terms_approx}), we return to the analysis of the pair of Hamiltonians 
(\ref{Hamiltonian}). The first observation is that a special form of the potentials $ U_{\pm}(y) $ in the  Schr{\"o}dinger Hamiltonians $ \mathcal{H}_{\pm} $ 
enables their factorization into two 
first-order operators as follows:
\beq
\label{factorization}
\mathcal{H}_{-} = \mathcal{A}^{+} \mathcal{A}, \; \; \; 
\mathcal{H}_{+} = \mathcal{A} \mathcal{A}^{+}
\eeq
where
\beq
\label{A}
\mathcal{A} = \frac{h^2}{\sqrt{2}}  \frac{\partial}{\partial y} +   \mathcal{W}(y) , \; \; \; 
\mathcal{A}^{+} = -  \frac{h^2}{\sqrt{2}}  \frac{\partial}{\partial y} +   \mathcal{W}(y) , \; \; \; 
 \mathcal{W}(y) :=   \frac{1}{\sqrt{2}}\frac{\partial V}{\partial y} 
\eeq 
The factorization properties (\ref{factorization}) are well known within supersymmetric quantum mechanics (SUSY QM) of Witten \cite{Witten_SUSY}.
The reason that the same mathematical construction arises in our problem is that  
the Schr{\"o}dinger equation (\ref{SE}) for the FPE possesses hidden supersymmetry (SUSY) \cite{Brown}, that makes it mathematically identical to the Euclidean supersymmetric quantum mechanics (SUSY QM), with $ h^2 $ playing the role of the Planck constant $ \hbar $. Using the standard nomenclature in the literature, we will refer to $  \mathcal{W}(y) :=   \frac{1}{\sqrt{2}}\frac{\partial V}{\partial y} $ as the \emph{superpotential}.

The factorization property (\ref{factorization}) comes very handy to find a zero-energy eigenstate $ \Psi_0 (y) = \Psi_0^{-} (y) $ of $ \mathcal{H}_{-} $ with $ E_{0}^{-} = 0 $, as it enables replacing a second-order ODE $ \mathcal{H}_{-}  \Psi_0  = 0 $ by a first order equation $ \mathcal{A} \Psi_0  = 0 $, or more explicitly,
\beq
\label{first_order_equation}
 \frac{h^2}{\sqrt{2}}  \frac{\partial  \Psi_0(y) }{\partial y} +   \mathcal{W}(y)  \Psi_0 (y) = 0
 \eeq
 The conventional use of this equation is to integrate it to obtain an expression of the zero-energy state $ \Psi_0 (y) $ in terms of the superpotential 
 $ \mathcal{W}(y) $:
 \beq
 \label{Psi_from_W}
 \Psi_0 (y)  = C e^{- \frac{\sqrt{2}}{h^2} \int_{0}^{y} \mathcal{W}(z) dz } = C e^{- \frac{V(y)}{h^2}  } 
 \eeq
 where $ C $ is a normalization constant.
 On the other hand, Eq.(\ref{first_order_equation}) can also be used in reverse in order to express the superpotential $ \mathcal{W}(y) $ in terms of $ \Psi_0^{-} (y) $:
 \beq
 \label{W_from_Psi}
 \mathcal{W}(y)  =  - \frac{h^2}{\sqrt{2}} \frac{d}{dy} \log \Psi_0 (y)
 \eeq   
 Therefore, if a zero-energy wave function $ \Psi_0 (y) = \Psi_0^{-} (y) $ of $ \mathcal{H}_{-} $ is known, the superpotential $ \mathcal{W}(y) $ is fixed in terms of this function. Moreover, as  $ \mathcal{W}(y) :=   \frac{1}{\sqrt{2}}\frac{\partial V}{\partial y} $, we also have an expression for the FPE potential $ V(y) $ in terms of $ \Psi_0 (y) $:
\beq
\label{V_from_Psi}
V(y) =  - h^2 \log \Psi_0 (y) + V_0
\eeq
where a constant $ V_0 $ is chosen such that the minimum value of $ V(y) $ is zero: $ V(y_{\star}) = 0 $. As in quantum mechanics a ground state wave function $ \Psi_0 (y) $ is always strictly positive \cite{Landau_QM}, Eq.(\ref{V_from_Psi}) is well-defined as long as $ \Psi_0 (y) $ is a valid ground state function of a quantum mechanical system. The representation (\ref{W_from_Psi}) is well known in the literature on SUSY QM, see e.g. \cite{SUSY_tunneling_asymmetric}. 

In our problem, the utility of the representation (\ref{V_from_Psi}) can be immediately seen by    
substituting this expression into Eq.(\ref{two_terms_approx}). This produces the following relation 
\beq
\label{two_terms_QM}
p(y,t | y_0) =   \left[ \Psi_0 (y) \right]^2 + 
 \frac{ \Psi_1^{-}(y_0)}{ \Psi_0 (y_0)} e^{- \frac{ t  \Delta E}{h^2}}  \Psi_0 (y) \Psi_1^{-}(y) + \ldots  
\eeq
This relation is also well-known in the literature, see e.g. \cite{vanKampen}. It shows that the stationary distribution $ p_s(y) $ of $ y_t $ as $ t \rightarrow \infty $ is given by the square of the ground-state wave function $  \Psi_0 (y) $: $  p_s(y) = \left[ \Psi_0 (y) \right]^2 $, in a very similar way to a probabilistic interpretation of quantum mechanics. The stationary density $ p_s(y) $ is normalized to one as long as the  $\Psi_0 (y) $ is squared-normalized. Moreover, due to orthogonality of wave functions $ \Psi_0(y) $ and $ \Psi_{1}^{-}(y) $, the transition density defined by Eq.(\ref{two_terms_QM}) is automatically normalized to one for arbitrary times $ t $. 

Traditionally, Eq.(\ref{two_terms_QM}) is used from the left to the right, 
implying that a solution of the FPE equation (\ref{FPE}) can be represented as in (\ref{two_terms_QM}), where the wave functions $  \Psi_0 (y),  \Psi_1^{-}(y) $ 
have to be computed by solving  the Schr{\"o}dinger equation with a given Hamiltonian $ \mathcal{H}_{-} $. In contrast, here we want to use it by reading from the \emph{right} to the \emph{left}, by \emph{defining} the FPE transition density  $ p(y,t | y_0) $ in terms of a given ground-state wave function $ \Psi_0 (y) $ viewed as a main modeling primitive, with higher-energy wave functions  $ \Psi_1^{-} (y) $ etc. derived from the same  $ \Psi_0 (y) $  as will be explained in details below.   

This implies, in particular, that if our only goal is to model a stationary distribution $ p_s(y) $ of $ y_t $ as $ t \rightarrow \infty $, it can be achieved by 
directly specifying a  quantum mechanical ground state wave function $\Psi_0 (y) $. A trial ground state wave function $ \Psi_0 (y) $ can be constructed as a parametric function $ \Psi_{\theta} (y) $, with a functional form and parameters $ \theta $ chosen such that  $ \Psi_{\theta} (y) $ has certain desired properties (such as a shape, symmetries, the local and global behavior at $ y \rightarrow \infty $, etc.).  Once a parametric function $ \Psi_{\theta} (y) $  is specified, as $  p_s(y) = \left[ \Psi_{\theta} (y) \right]^2 $, defining a stationary distribution thus requires zero extra parameters or calculations.
 
Note that when viewed on its own, the idea of using the relation $ p_s(y)  = \left[ \Psi_{\theta} (y) \right]^2$ as a way to model a stationary distribution of the FPE dynamics may appear a kind of trivial, as any stationary probability density can be represented as a square of another function. The real value of using the ground state wave function $  \Psi_0 (y) = \Psi_{\theta}(y) $ as a prime modeling primitive becomes apparent when it is used jointly with the relation (\ref{V_from_Psi}) that defines the potential $ V(y) $ in terms of  $  \Psi_0 (y) $. Therefore, a parametric specification $  \Psi_0 (y) = \Psi_{\theta}(y) $ translates into a parametric choice 
for $ V(y) = V_{\theta}(y) $ according to Eq.(\ref{V_from_Psi}). Once the potential $ V_{\theta}(y) $ or the superpotential $ \mathcal{W}_{\theta}(y) =  
\frac{1}{\sqrt{2}}\frac{\partial V_{\theta}}{\partial y} $ are defined,  
the first excited state $ \Psi_1^{-}(y) $ (and higher states) can now be obtained as a solution of the the Schr{\"o}dinger equation (\ref{SE}). In this equation, 
the Hamiltonian $\mathcal{H}_{-} = \mathcal{A}^{+} \mathcal{A}$,  with $ \mathcal{A} $ given by Eq.(\ref{A}) where the superpotential $ \mathcal{W}_{\theta}(y) $ is defined by Eq.(\ref{W_from_Psi}). This procedure fixes all time-dependent terms in Eq.(\ref{two_terms_QM}) in terms of the only model input $  \Psi_0 (y) = \Psi_{\theta}(y)$.

This suggests a way of constructing dynamic 
probability distributions with desirable properties, whose parameters could be interpreted in terms of parameters $ \theta $ that specify the ground-state wave function $\Psi_{\theta} (y) $. Note that while defining a superpotential $ \mathcal{W}(y) $ in terms of a properly chosen parameterized ground state wave function was previously used to compute tunneling rates with bistable potentials in the context of SUSY QM, see e.g. \cite{SUSY_tunneling_asymmetric}, here we propose a similar idea to model non-linear \emph{stochastic processes} in a tractable way, by specifying them in terms of a parametrized ground state wave function (or a 'half-probability distribution') $ \Psi_{\theta} (y) $. As suggested by  Eq.(\ref{two_terms_QM}), if our only objective is a limiting stationary distribution  $ p_s(y) $ of $ y_t $ as $ t \rightarrow \infty $, 
no further work is needed, and $ p_s(y)  = \left[ \Psi_{\theta} (y) \right]^2$. 
 
To summarize, the representation (\ref{two_terms_QM}) of a time-dependent transition density of the FPE (\ref{FPE}) provides a new tractable way to model  
non-linear Langevin dynamics (\ref{Langevin_return_1M}), where the interaction potential $ V(y) $ is defined as in Eq.(\ref{V_from_Psi}). 
With this approach, calculation of a stationary density $ p_s(y) $ requires no calculation at all, as $ p_s(y)  = \left[ \Psi_0 (y) \right]^2$. 
This can be directly computed once a parametrized trial function $  \Psi_0 (y) = \Psi_{\theta}(y)$ is specified.
To compute a time-dependent pre-asymptotic behavior due to the second term in (\ref{two_terms_QM}), the only additional step needed is 
a calculation of the first excited state $ \Psi_{1}^{-} $ and the energy $ E_{1}^{-} $ of the Hamiltonian $ \mathcal{H}_{-} $. In the next section, we will consider 
a simple model of a ground-state wave function  $ \Psi_0 (y) $ where such calculations become quite tractable. 

\subsection{NES: A Log-Gaussian Mixture double well potential} 
\label{sect_Gaussian_mixture_WF}

\begin{figure}[ht]
\begin{center}
\includegraphics[
width=170mm,
height=110mm]{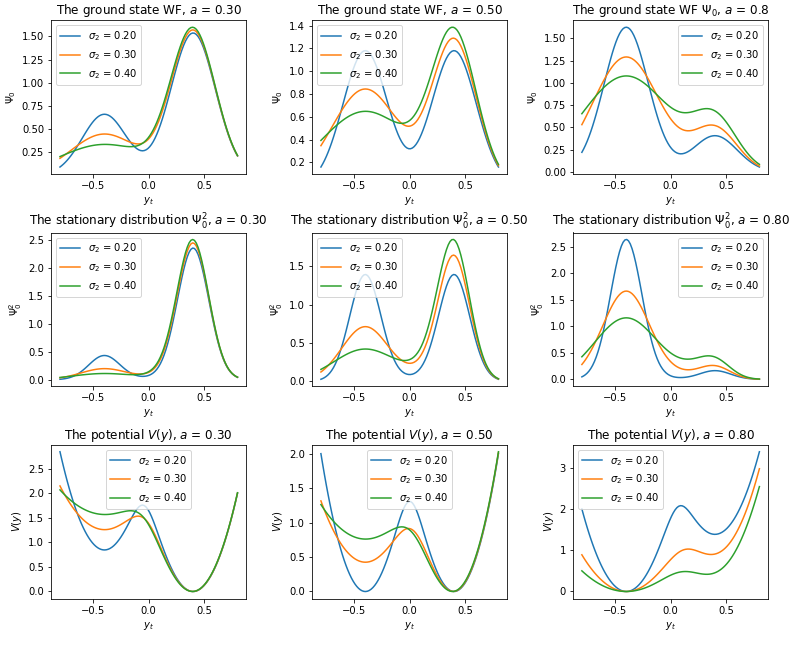}
\caption{The ground state wave function $ \Psi_0(y) $, the stationary distribution $ \Psi_0^2(y) $ and the Langevin potential (\ref{pot_explicit}) as a function of the market return $ y_t $, for a few values of the asymmetry parameter $ a $, and with different values of parameters $ \sigma_2 $, with fixed values $ \sigma_1 = 0.2,
\mu_1 = 0.4, \mu_2 = - 0.4 $, $ T = 1 $.  Graphs on the left describe possible shapes for a 'normal' market regime, where the right minimum is a global minimum. 
Graphs on the right illustrate a crisis-like regime,  when the left minimum becomes  a global minimum. Graphs in the middle column describe intermediate scenarios. 
In particular, when $ a = 0.5 $ and $ \sigma_1 = \sigma_2 $, the resulting potential shown in the blue line is symmetric, and can describe a scenario with a spontaneous symmetry breaking.}
\label{fig_IQED_WF_and_potentials}
\end{center}
\end{figure}

 
 To produce an 
 interesting and tractable formulation, we consider a two-component
 Gaussian mixture (GM) as a model of $ \Psi_0(y) $, with means $ \mu_1 T $ and $ \mu_2 T < \mu_1 T$ and a mixture coefficient $ 0 \leq a \leq 1 $,
with a normalization parameter $ C $ chosen such that $ \Psi_0(y) $
 is squared-integrable with $ \int \Psi_0^2(y) dy = 1 $: 
 \beq
 \label{Psi_0}
 \Psi_0(y) = C \left[ (1-a) \phi(y | \mu_1 T, \sigma_1^2 T) + a  \phi(y | \mu_2 T, \sigma_2^2 T) \right], \; \; \;  \int \Psi_0^2(y) dy = 1
 \eeq
 Such a combination of Gaussian densities often provides a good representation of a ground state wave function of a particle in a double well potential.
 Double well potentials play a special role in statistical physics and quantum mechanics, and are often used to model tunneling phenomena, see e.g. 
 \cite{Landau_QM} or \cite{Zinn-Justin-QFT}. In particular, a symmetric double well is described by a symmetric version of (\ref{Psi_0}) with $ a = 1/2 $ and $ \mu_1 = - \mu_2, \, \sigma_1 = \sigma_2 $.
 For other choices of model parameters, the GM model (\ref{Psi_0}) can fit a variety of shapes including both a unimodal and bimodal shapes. 
 For a later use, 
 it is convenient to represent the WF (\ref{Psi_0}) as a Gaussian pdf multiplied by a factor that  asymptotically approaches a constant:
 \beq
\label{Psi_0_new}
\Psi_{0} (y) = 
\left\{ \begin{array}{ll}
a C   \phi(y | \mu_2 T, \sigma_2^2 T) \eta(y)
 & \text{if} \;  \sigma_2 >  \sigma_1  \\
(1-a)  C   \phi(y | \mu_1 T, \sigma_1^2 T) \eta (y)  
   & \text{if} \; \sigma_2  < \sigma_1   \\ 
(1-a)  C   \phi(y | \mu_1 T, \sigma_1^2 T) \eta (y)  
 & \text{if} \;  \sigma_2 =  \sigma_1, \, y \geq 0  \\
  a C   \phi(y | \mu_2 T, \sigma_2^2 T) \eta(y)
   & \text{if} \; \sigma_2  = \sigma_1, \, y < 0   \\  
\end{array} \right.
\eeq
where 
function $ \eta (y) $ is defined as follows:
 \beq
\label{eta_fun}
\eta (y) = 
\left\{ \begin{array}{ll}
 1 + \frac{1-a}{a} \exp\left\{ \log \phi(y | \mu_1 T, \sigma_1^2 T) - \log \phi(y | \mu_2 T, \sigma_2^2 T)    \right\}    & \text{if} \; \sigma_2 > \sigma_1   \\
 1 + \frac{a}{1-a} \exp\left\{  \log \phi(y | \mu_2 T, \sigma_2^2 T)  -\log \phi(y | \mu_1 T, \sigma_1^2 T)  \right\}    & \text{if} \; \sigma_2 < \sigma_1   \\
  1 + \frac{a}{1-a} \exp\left\{  \log \phi(y | \mu_2 T, \sigma_1^2 T)  -\log \phi(y | \mu_1 T, \sigma_1^2 T)  \right\}    & \text{if} \; \sigma_2 =  \sigma_1, \,  y \geq 0 \\
  1 + \frac{1-a}{a} \exp\left\{ \log \phi(y | \mu_1 T, \sigma_1^2 T) - \log \phi(y | \mu_2 T, \sigma_1^2 T)    \right\}    & \text{if} \; \sigma_2  = \sigma_1, \, y < 0   
 \end{array} \right.
\eeq
 Viewing (\ref{Psi_0}) as an \emph{exact}, rather than an approximate ground state wave function, we can plug it into Eq.(\ref{V_from_Psi}) to find the explicit form of the Langevin potential $ V(y) $:
 \beq
 \label{pot_explicit}
  V (y)  = - h^2 \log  \left[ (1-a) \phi(y | \mu_1 T, \sigma_1^2 T) + a  \phi(y | \mu_2 T, \sigma_2^2 T) \right] - h^2 \log C + V_0
  \eeq
 where again $ V_0 $ is a constant chosen such that the minimum value of $ V(y) $ is zero. Examples of trial ground state WFs $ \Psi_0 $ and resulting potentials $ V(y) $ 
 are shown in Fig.~\ref{fig_IQED_WF_and_potentials}.
 
 While this expression produces a non-linear behavior for small positive or negative values of $ y $, its limiting behavior at $ y \rightarrow \pm \infty $ is rather simple and coincides with a harmonic (quadratic) potential:
 \beq
 \left. V(y) \right|_{y \rightarrow - \infty} = h^2 \frac{(y-\mu_2 T)^2}{2 \sigma_2^2 T}, \; \; \; 
 \left. V(y) \right|_{y \rightarrow  \infty} = h^2 \frac{(y-\mu_1 T)^2}{2 \sigma_1^2 T} \; \; \; \; (\mu_2 < \mu_1) 
 \eeq  
The fact that the limiting behavior of the potential coincides with a harmonic potential as $ y \rightarrow \pm \infty $ means that  in this asymptotic regime the model behavior is described by a harmonic oscillator, and thus is fully analytically tractable.

 As Gaussian mixtures are known to be universal approximations for an arbitrary non-negative functions given enough components, this implies that an \emph{arbitrary} potential that asymptotically coincides with a harmonic oscillator potential can be represented as a negative logarithm of a Gaussian mixture. We can refer to such class of potentials as Log-Gaussian Mixture (LGM) potentials. In our particular case, we restrict ourselves to a two-component LGM potential.\footnote{A requirement of an asymptotic harmonic oscillator behavior could be seen as a potential limitation for the LGM class of potentials, as many interesting potentials
 have a different asymptotic behavior. To this point, we can note that an onset of such a quadratic regime can always be pushed further away by a proper rescaling of the coordinate, while for small or moderate values of a new rescaled argument, the dynamics can still be arbitrarily non-linear, and driven by the number of Gaussian components and their parameters.}

\begin{figure}[ht]
\begin{center}
\includegraphics[
width=180mm,
height=85mm]{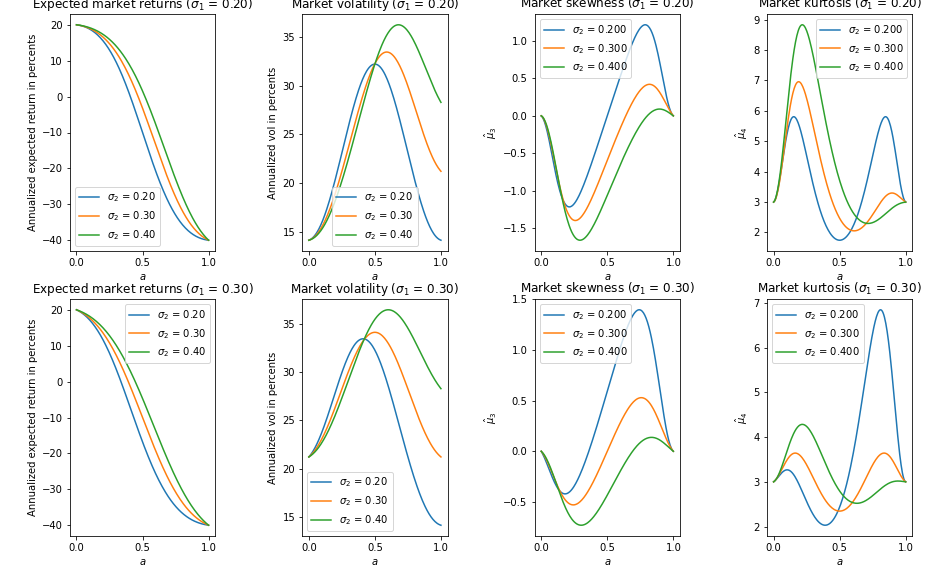}
\caption{Higher moments of the equilibrium distributions $ \Psi_0^2(y) $ as functions of the asymmetry parameter $ a $, and with different values of parameters $ \sigma_2 $.}
\label{fig_higher_moments}
\end{center}
\end{figure}

 Note that  while  $ \Psi_0(y) $ is proportional to a \emph{two}-component Gaussian mixture, its square is proportional to a \emph{three}-component Gaussian mixture:
 \bea
 \label{three_component_Psi_2}
 \Psi_0^2 (y) &=& \frac{C^2}{2 \sqrt{\pi T}} \left[ \frac{(1-a)^2}{\sigma_1} \phi \left(y | \mu_1 T, \frac{\sigma_1^2}{2} T\right) + 
  \frac{a^2}{\sigma_2} \phi \left(y | \mu_2 T, \frac{\sigma_2^2}{2}  T\right)  \right. \nonumber \\
 & + & \left.
  \frac{ 2 a(1-a)  }{\sqrt{(\sigma_1^2 + \sigma_2^2)/2}} e^{ - \frac{ (\mu_1 - \mu_2)^2 T}{2(\sigma_1^2 + \sigma_2^2)} } 
   \phi \left(y | \mu_3 T, \frac{\sigma_3^2}{2} T \right) \right]
 \eea
  where the additional third Gaussian component has the following mean and variance:
 \beq
 \label{third_component}  
 \mu_3 := \frac{\mu_1 \sigma_2^2 + \mu_2 \sigma_1^2}{\sigma_1^2 + \sigma_2^2}, \; \; \; 
  \frac{\sigma_3^2}{2}  =   \frac{ \sigma_1^2 \sigma_2^2  }{\sigma_1^2 + \sigma_2^2}
 \eeq
 The normalization condition thus fixes the value of the constant $ C $ as follows:
 \beq
 \label{C}
 C^2 =  \frac{2 \sqrt{\pi T}}{ \Omega}, \; \; \; \text{where} \; \; \; \Omega =  \frac{(1-a)^2}{\sigma_1} + 
  \frac{a^2}{\sigma_2}  + 
  \frac{ 2 a(1-a)}{\sqrt{(\sigma_1^2 + \sigma_2^2)/2}} e^{ - \frac{ (\mu_1 - \mu_2)^2 T}{2(\sigma_1^2 + \sigma_2^2)} } 
  \eeq 
Recalling the relation  $ p_s(y)  = \left[ \Psi_0 (y) \right]^2$ and introducing weights
\beq
\label{GM_weights}
\omega_1 := \frac{(1-a)^2}{\sigma_1 \Omega}, \; \; \; 
\omega_2 := \frac{a^2}{\sigma_2 \Omega}, \; \; \;
 \omega_3 :=   \frac{2 a(1-a)}{ \Omega \sqrt{(\sigma_1^2 + \sigma_2^2)/2}} e^{ - \frac{ (\mu_1 - \mu_2)^2 T}{2(\sigma_1^2 + \sigma_2^2)} }, \; \; \; 
 \sum_{i=1}^{3} \omega_i = 1  
\eeq
we obtain the following three-component Gaussian mixture model for the stationary distribution $ p_s(y) = \Psi_0^2(y) $:
 \beq
\label{GM_3_comp}
 p_s(y)  =  \sum_{k=1}^{3} \omega_k \phi \left(y | \mu_k T, \frac{\sigma_k^2}{2} T \right) 
 \eeq
Note that while a priori defining a three-component Gaussian mixture model requires eight parameters, our approach fixes all parameters of the Gaussian mixture 
(\ref{GM_3_comp}) in terms of only five original model parameters $ \mu_1, \mu_2, \sigma_1, \sigma_2, a $, or rather four parameters if we set $ \mu_1 = \mu_2 $ as we will do in most of examples below. Interestingly, the fifth model parameter $ h $ drops off in
the expression for the stationary density $ p_s(y) $, and only appears when treating non-stationary effects due to the second term in Eq.(\ref{two_terms_QM}).
In addition, as we will show next, this parameter also controls intensity of transitions between minima of the potential (\ref{pot_explicit}).
The behavior of moments of the equilibrium distribution is shown in Fig.~\ref{fig_higher_moments}.

\subsection{Dynamics with a bistable potential: metastability and instantons}
\label{Instantons_and_Kramers}

As illustrated in Fig.~\ref{fig_IQED_WF_and_potentials}, the NES model may produce a variety of equilibrium distributions including both 
unimodal and bimodal shapes, where transitions between different shapes are controlled by model parameters. Informally, one can expect that unimodal distributions would be typical for benign markets, while bimodal distributions may arise when a market is distressed or in a crisis. 
In this section, we focus on a model regime when parameters are such that a ground state $ \Psi_0(y) $ is a bimodal function, corresponding to the market in a crisis or a severe distress. However, the final formulae that we will carry to a final model formulation will be applicable for any choice of $ \Psi_0 $, 
whether it is unimodal or bimodal, or equivalently for any specification a LGM potential, whether it has a single minimum or two local minima.  Formulas that specifically assume a particular choice of the potential will be separately mentioned below.

Assume now that the market is in a regime when $ \Psi_0(y) $ has two maxima separated by a local  
minimum, see Fig.~\ref{fig_IQED_WF_and_potentials}. As the logarithmic function is monotonic, Eq.(\ref{V_from_Psi}) then implies a bimodal Langevin potential $ V(y) $, with a local 
maximum at $ y_m $ separating two local minima $ y_{\star} $ and $ y^{\star} $ (the latter coincide with maxima of $ \Psi_0(y) $).
Here we assume that $ y_{\star} < y_m < y^{\star} $, and that $ V(y_{\star}) \leq   V(y^{\star}) $, so that $ y_{\star} $ and $ y^{\star} $ are the global and local minima, respectively. 
The height of a potential barrier faced by a particle initially located at $ y^{\star} $ (i.e. in a local minimum) is 
$ \Delta V := V(y_m) - V(y^{\star}) $. Such potentials lead to metastability, where a particle initially located near a local minimum $ y^{\star} $ hopes over the barrier 
$ \Delta V $ as a result of thermal fluctuations. Following physics conventions, we will occasionally refer to states $ y_{\star} $ and $ y^{\star} $ as the true and metastable vacua, respectively.  
Using this nomenclature, a situation with a potential with a single minimum can be described as a theory with a single 
stable vacuum. 
  

Using the FPE approach, an escape rate from a metastable minimum of a potential $ V(y) $ can be found as the inverse of a mean passage time $ \mathcal{T}(y) $ to exit the interval $ y = [y_a, y_b] $, viewed as a function of the \emph{backward} (time-zero) variable $ y = y_0 $, assuming that $ y_a < y_0 < y_b $. In our case, we can set $ y_b = \infty $ and $ y_a = y_{\star} $, with $ y_0 > y_{\star} $, so that $ [y_a, y_b] = [y_{\star}, \infty] $ and  $ y_{\star} \leq y_0 \leq \infty $. For a time-homogeneous process,  the mean passage time $ \mathcal{T}(y) $ satisfies the following backward PDE (see e.g. \cite{Gardiner}):
\beq
\label{Mean_T_PDE}
- V'(y) \frac{ \partial \mathcal{T}(y)}{ \partial y} + \frac{h^2}{2} \frac{\partial^2 \mathcal{T}(y)}{\partial y^2} = -1
\eeq
with boundary conditions $  \mathcal{T}(y_{\star}) = 0 $ and $  \mathcal{T}'(\infty) = 0 $. 
This equation can be solved by multiplying by an integrating factor $ e^{-2 V(y)/h^2} $, and then twice integrating the resulting expression using the boundary conditions. This produces the following well-known relation for the mean passage time  $ \mathcal{T}(y) $:
\beq
\label{Tau}
\mathcal{T}(y_0) = \frac{2}{h^2} \int_{y_{\star}}^{y_0} dy e^{ \frac{2 V(y)}{h^2}} \int_{y}^{\infty} dz  e^{ - \frac{2 V(z)}{h^2}}
\eeq
This is a general expression valid for any potential $ V(y) $ that is sufficiently well-behaved at $ y \rightarrow \pm \infty $. In particular, it applies to both cases of a potential with a single minimum or a potential with two local minima separated by a maximum. 
For the former case, $ y_{\star} $ is not a position of a global minimum of the potential, but rather a threshold value of log-return that signals a crisis regime of the market. For the latter case, Eq.(\ref{Tau}) assumes that the 'particle' is initially placed in the right well corresponding to a local minimum $ y^{\star} $, and eventually escapes to the left well with a global minimum of the potential at $ y = y_{\star} $.  

For a potential with two minima, the mean passage time can be calculated using quadratic expansions of  $ V(y) $, where the first and second integrals in 
Eq.(\ref{Tau}) are computed using a saddle point approximation with quadratic expansions around, respectively, a maximum 
$ y_m $ and the local minimum  $ y^{\star} $.  Such an approximation is justified when $ \Delta V/h^2  \gg 1 $, i.e. when the barrier is high,
while the initial position $ y_0 $ is near the local minimum $ y^{\star}$. This produces the celebrated 
Kramers escape rate formula for the escape intensity $ \lambda = 1/\mathcal{T} $ (see e.g. \cite{Gardiner} or \cite{Hanggi_1986}):
\beq
\label{Kramers_rate_main}
\lambda = \frac{\sqrt{ V''(y^{\star}) \left| V''(y_m) \right| }}{2 \pi} \exp \left[ - \frac{2 \Delta V }{h^2}\right]
\eeq
The same result (\ref{Kramers_rate_main}) can also be obtained using alternative methods.
In particular, with the  Schr{\"o}dinger equation (\ref{SE}), the Kramers escape rate $ \lambda $ can be calculated as $ \lambda =  \Delta E/ h^2 $
where $ \Delta E = E_{1}^{-} - E_0  =   E_{1}^{-} $ is the energy splitting between a ground state and a first excited state \cite{Brown}. Within a path integral approach, 
the energy splitting $ \Delta E $ can be obtained as a contribution to the path integral due to \emph{instantons} - saddle-point solutions of dynamics obtained in a weak noise (quasi-classical) limit  $ h \rightarrow 0 $ of the Langevin equation (\ref{Langevin_return_1M}), where the potential is \emph{inverted}. The instanton equation reads (see Appendix A for more details)
\beq
\label{instanton}
\frac{d y_t}{d t} = \frac{ \partial V (y_t)}{\partial y_t}
\eeq   
While instantons produce the exponential term in (\ref{Kramers_rate_main}),
the pre-exponential factor is obtained from thermal (or quantum) corrections to an instanton contribution to a transition probability for 
 a bimodal or metastable potential $ V(y) $, see e.g. \cite{HD_QED}.
Note that the Kramers escape rate (\ref{Kramers_rate_main}) is non-analytic in the 'Planck constant' $ h^2 $, i.e. it does not have a Taylor expansion around the value $ h^2 = 0 $. Instantons arising in statistical and 
quantum mechanics or in quantum field theory produce a similar non-analyticity of transition amplitudes in the Planck constant $ \hbar $ or in coupling constants driving non-linearities, see e.g. \cite{Coleman_book}. This means that instantons are \emph{non-perturbative} solutions that can not be found to any order of a perturbation theory in $ \hbar $. On the other hand, using methods of supersymmetry for analysis of metastability in the Fokker-Planck dynamics with a bistable potential, the same results as obtained from instanton calculus can also be obtained using a pure quantum mechanical approach, as was shown in \cite{Brown}.
  

With our approach where the ground state wave function is the main modeling primitive, we can express $ \mathcal{T}(y) $ in Eq.(\ref{Tau}) by substituting there 
Eq.(\ref{V_from_Psi}). This gives
 \beq
\label{Tau_from_Psi}
\mathcal{T}(y_0) = \frac{2}{h^2} \int_{y_{\star}}^{y_0} \frac{ dy}{ \Psi_0^2(y)} \int_{y}^{\infty} dz  \Psi_0^2(z) 
\eeq
Note that unlike Eq.(\ref{Kramers_rate_main}) which only applies for a bistable potential, the last expression (\ref{Tau_from_Psi}) is applicable for either a bistable potential or a potential with a single minimum. In addition, numerical integration in Eq.(\ref{Tau_from_Psi}) can be done with arbitrary precision, unlike 
the Kramers relation (\ref{Kramers_rate_main}) which is based on the saddle point approximation and the presence of two minima of the potential. Therefore, we will use Eq.(\ref{Tau_from_Psi}) instead of  (\ref{Kramers_rate_main}) in a practical implementation of the model to be presented below.  

Interestingly, the last relation (\ref{Tau_from_Psi}) suggests that when $ \Psi_0^2 $ is provided as an input that does not depend explicitly on parameter $ h $, then the whole expression depends on $ h $ as $ h^{-2} $, which means that the escape rate $ \lambda = 1/\mathcal{T} $ is proportional to the 'Planck constant' $ h^2 $. We will return to this remark below. 

\begin{figure}[ht]
\begin{center}
\includegraphics[
width=80mm,
height=55mm]{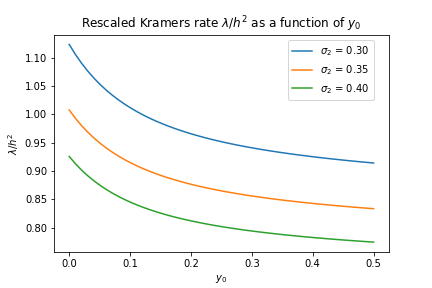}
\caption{Rescaled Kramers rate $ \lambda / h^2  = 1/ ( \mathcal{T} h^2) $ as a function of  the initial position $ y_0 $ computed by a numerical integration in Eq.(\ref{Tau_from_Psi}).  Model parameters used here are $ \mu_1 = - \mu_2 = 0.2, \, a = 0.2, \, \sigma_1 = 0.2 $.} 
\label{fig_rescaled_Kramers_vs_y0}
\end{center}
\end{figure}

While Eq.(\ref{Tau_from_Psi}) is general and applies for any ground state WF $ \Psi_0 $, it can be approximately calculated analytically for a bistable potential,
producing corrections to the Kramers relation (\ref{Kramers_rate_main}).  To show this, 
we use the expression for $ p_s(y) = \left[ \Psi_0(y) \right]^2 $ given for the trial function (\ref{Psi_0}) by Eq.(\ref{GM_3_comp}), and compute the inner integral
in (\ref{Tau_from_Psi}) explicitly:
\beq
\label{explicit_inner}
 \int_{y}^{\infty} dz  \Psi_0^2(z) =  \sum_{k=1}^{3} \omega_k  \mathcal{N} \left(\frac{\sqrt{2} (\mu_k T - y)}{\sigma_k \sqrt{T} } \right) 
 \eeq
where $ \mathcal{N}(\cdot) $ stands for the cumulative normal distribution. We approximate the outer integral in (\ref{Tau_from_Psi}) by reverting to its previous expression in Eq.(\ref{Tau}), and expanding the integrand around its maximum at 
$ y_m $, while simultaneously replacing the integration limits by infinities. This gives 
\beq
\label{approximate_T}
\mathcal{T}(y_0) = \frac{2}{h^2 C^2} e^{\frac{2 V(y_m)}{h^2}}   \int_{-\infty}^{\infty}  e^{ - \frac{ (y-y_m)^2 }{2 \sigma_m^2} }   \int_{y}^{\infty} dz  \Psi_0^2(z)
\eeq
where 
\beq
\label{sigma_hat} 
\sigma_m^2 = \frac{h^2}{2 \left|  V''(y_m)  \right|}, \; \; \; \frac{1}{C^2} = \int_{-\infty}^{\infty} e^{ - \frac{2 V(y)}{h^2}} dy \simeq \sqrt{\frac{2 \pi h^2}{2 V''(y^{\star})} }
e^{ - \frac{2 V(y^{\star})}{h^2}} 
\eeq
Here we replaced the exact expression (\ref{C}) for the normalization constant $ C $ by its Gaussian approximation obtained using a quadratic expansion 
of $ V(y) $ around its local minimum $ y^{\star} $. 
Using Eq.(\ref{explicit_inner}), the integral in Eq.(\ref{approximate_T}) can be computed analytically using the formula\footnote{See e.g. https://en.wikipedia.org/wiki/List$\_$of$\_$integrals$\_$of$\_$Gaussian$\_$functions\#CITEREFOwen1980.} 
\beq
\label{integral_of_cum_norm}
\int_{-\infty}^{\infty} \phi(x) \mathcal{N} \left( a + b x  \right)  dx =  \mathcal{N} \left(  \frac{a}{\sqrt{1 + b^2 } } \right)
\eeq
This produces the final analytical approximation for the mean passage time, which applies when an initial position $ y_0 $ is near the local minimum $ y^{\star} $:
\beq
\label{Tau_final}
\mathcal{T} =  \frac{2 \pi}{\sqrt{ V''(y^{\star}) \left| V''(y_m) \right| }}  e^{ \frac{2 \Delta V}{h^2}}
  \sum_{k=1}^{3} \omega_k \mathcal{N} \left( \frac{\sqrt{2} (\mu_k T - y_m) }{ \sqrt{(\sigma_k^2 + 2 \sigma_m^2) T}} \right)  
\eeq
When the arguments of  cumulative normal functions in this relation can be approximated by infinities (which is formally obtained in the limit the limit $ \mu_k - y_m \rightarrow \infty $ or alternatively $ \sigma_k, \sigma_m \rightarrow 0 $), this expression produces the same formula (\ref{Kramers_rate_main}) for the escape rate $ \lambda = 1/ \mathcal{T} $. Note that a dependence on the initial position $ y_0 $ is neglected in Eq.(\ref{Tau_final}) within the saddle-point approximation. 
A direct numerical integration in Eq.(\ref{Tau_from_Psi}) (which is used for a practical implementation, as was mentioned above) shows that the saddle-point approximation becomes less accurate for smaller initial values $ y_0 $, see Fig.~\ref{fig_rescaled_Kramers_vs_y0}. 

It is instructive to check how the mean passage time (\ref{Tau_final}) depends on the 'Planck constant' $ h^2 $. Note that $ \sigma_m $ defined in Eq.(\ref{sigma_hat}) is actually independent of $ h $ as implied by (\ref{V_from_Psi}). As other parameters entering the sum do not depend on $ h $ while $ V''(y) \sim h^2 $ (see Eq.(\ref{pot_explicit})),  we have  $ \mathcal{T} \sim h^{-2} $. As was mentioned above, the same parametric dependence can also be seen directly from Eq.(\ref{Tau_from_Psi}).

Therefore, parameter $ h^2 $ controls the escape rate to a stable vacuum starting with a metastable vacuum. On the other hand, note that  
$ h^2 $ does \emph{not} appear in the trial wave function (\ref{Psi_0}) or the stationary distribution (\ref{GM_3_comp}). This does not mean, of course, that  properties of a stationary state are not affected by the volatility parameter $ h $, but rather means that with our method of a direct parameterization of the trial 
wave function (\ref{Psi_0}), its impact on an equilibrium distribution is already incorporated in other model parameters.\footnote{This can be also be simply seen in the original FPE equation (\ref{FPE}): if we re-define the potential $ V(y) = - h^2 U(y) $ where $ U(y) $ is another potential, and then look for a stationary solution,
the constant $ h^2 $ drops out in the resulting stationary distribution formula when expressed in terms of $ U(x) $. It does not imply, of course, that $ h^2 $ magically disappears from the problem, but simply means that for a stationary distribution, it can be formally absorbed into parameters defining the new potential $U(y) $.}   
With this parametrization, the Kramers escape rate $ \lambda $ is equal to $ h^2 $ times a \emph{fixed} function of other model parameters, which is illustrated in Fig.~\ref{fig_rescaled_Kramers_vs_y0}. 
As will be shown in the next section, parameter $ h^2 $ also appears in pre-asymptotic corrections to the asymptotic (stationary) behavior of the model, and is therefore related to the speed of a relaxation to such an equilibrium state.

\section{Computing dynamic corrections with SUSY}
\label{SUSY_higher_states}

\subsection{Supersymmetry and dynamics}
\label{sect_SUSY_and_dynamics}

So far, our discussion was centered around properties of a stationary distribution (\ref{Psi_0}) that corresponds to an asymptotic (at time $ t \rightarrow \infty $) behavior of the FP transition density $ p(y,t| y_0) $ in the Langevin model with the potential $ V(y) $ given in terms of a ground state WF $ \Psi_0(y) $ 
in Eq.(\ref{pot_explicit}). As was discussed above, using a parametric model $ \Psi_0(y) = \Psi_{\theta}(y) $ (where in our case $ \theta $ stands for all parameters entering Eq.(\ref{Psi_0})) is very convenient as it provides a direct answer to the problem of finding the asymptotic behavior of the model (assuming all parameters are kept fixed) as $ t \rightarrow \infty $. As can be seen from Eq.(\ref{two_terms_QM}), the equilibrium distribution $ p_s $ that describes this limit is 
given by  $ \Psi_0^2(y) $, so that no further work is needed to find it. 

Therefore, when the stationary distribution  $ \Psi_0^2(y) $ is an input, rather than an output in a model, the burden of computing the \emph{dynamics} is considerably reduced - all we need to find 
according to Eq.(\ref{two_terms_QM}) are higher-energy states $ \Psi_n^{-}(y) $ with $ n = 1, 2, \ldots $, and their energies $ E_{n}^{-} $ They should be computed as solution of the the Schr{\"o}dinger equation (\ref{SE}). In this equation, 
the Hamiltonian $\mathcal{H}_{-} = \mathcal{A}^{+} \mathcal{A}$,  with $ \mathcal{A} $ given by Eq.(\ref{A}) where the superpotential $ \mathcal{W}_{\theta}(y) $ is defined by Eq.(\ref{W_from_Psi}). 

This procedure fixes all time-dependent terms in Eq.(\ref{two_terms_QM}) in terms of the only model input $  \Psi_0 (y) = \Psi_{\theta}(y)$.
The challenging part is to actually compute these higher-energy states WFs  $ \Psi_n^{-}(y) $ with $ n = 1, 2, \ldots $. For some potential, the energy split 
$ \Delta E = E_{1}^{-} - E_0 $ can be very small, that may give rise to substantial challenges when computing it, along with computing WFs of excited states using various approximation methods. Furthermore, while many approximate methods of quantum mechanics work well for a ground state of a QM system, they become harder to use (or not usable at all) for higher states.

Supersymmetry (SUSY) is known to offer very efficient computational methods to solve such problem, which were applied to various problems in 
SUSY QM and classical stochastic systems, see e.g. \cite{SUSY_QM_Cooper} and \cite{Junker}, respectively.
Supersymmetry of the Hamiltonians $ \mathcal{H}_{\pm} $ implies that all eigenstates of $ \mathcal{H}_{-}  $ should be degenerate in energy with eigenstates of the SUSY partner Hamiltonian  $ \mathcal{H}_{+} $, except 
possibly for a 'vacuum' state with energy $ E_0^{-} = 0 $ \cite{Witten_SUSY}. Such zero-energy ground state would be unpaired, while all higher states would be doubly degenerate between the SUSY partner Hamiltonians $ \mathcal{H}_{\pm} $ according to the following relations (see e.g. \cite{SUSY_QM_Cooper} or \cite{Junker}, or Appendix B for a short summary):
\bea
\label{SUSY_relations_main}
&&  \mathcal{H}_{-} \Psi_0^{-} = \mathcal{A}  \Psi_0^{-} = 0, \; \;   E_0^{-} = 0 \nonumber \\
&& \Psi_{n+1}^{-} = \left(E_n^{+} \right)^{-1/2}
\mathcal{A}^{+} \Psi_{n}^{+}, \; \; \; 
\Psi_{n}^{+} = \left(E_{n+1}^{-} \right)^{-1/2}
\mathcal{A} \Psi_{n+1}^{-}, \; \; \; n = 0, 1, \ldots \\
&& E_{n+1}^{-} = E_{n}^{+}, \; \; \; n = 0, 1, \ldots 
\nonumber 
\eea
These relations assume that a zero-energy state $ \Psi_0^{-}  $ exists. In certain models, Hamiltonians $ \mathcal{H}_{\pm} $ can be supersymmetric, while 
a ground state  $ \Psi_0^{-}  $ has a non-zero energy $ E_0^{-} > 0 $.  Such scenarios correspond to a spontaneous breakdown of SUSY \cite{Witten_SUSY}.
For such models, supersymmetry is a symmetry of a Hamiltonian but not of a ground state of that Hamiltonian. On the other hand, an unbroken SUSY is characterized by the existence of a normalizable ground state $ \Psi_0 $ with strictly zero energy $ E_0 = 0 $.
For SUSY to be unbroken, the derivative of the superpotential 
$ V'(y) = \partial V/ \partial y $ should have different signs at $ y = \pm \infty $, which means that that it should have an odd number of zeros at real values 
of $ y $ \cite{Witten_SUSY}. 
In our problem, SUSY is \emph{not} spontaneously broken by design, as the ground state WF $ \Psi_0(y) $ defined in (\ref{Psi_0}) has a zero energy by construction, 
and the superpotential $ \mathcal{W}(y) $ in Eq.(\ref{W_from_Psi}) does have different signs at  $ y = \pm \infty $.

Therefore, we can use the SUSY relations to compute $ \Psi_1^{-} $ (and other higher states, if needed) as described above. 
The reason SUSY is helpful in practice is that it allows one to replace a hard problem of computing a first excited state for the Hamiltonian $ \mathcal{H}_{-} = \mathcal{A}^{+}  \mathcal{A} $ by a simpler problem of computing a ground state WF for its SUSY partner $ \mathcal{H}_{+} =   \mathcal{A} \mathcal{A}^{+} $. Once the ground state WF $ \Psi_0^{+} $ of the Hamiltonian  $ \mathcal{H}_{+} $
is computed using some approximation (e.g. a variational or a perturbation method),  the first excited state  $ \Psi_1^{-} $ of $ \mathcal{H}_{-} $ can be obtained 
according to the SUSY relations (\ref{SUSY_relations_main}) by applying the SUSY generator $ \mathcal{A}^{+} $ to  $ \Psi_0^{+} $, where the superpotential $ \mathcal{W}_{\theta}(y) $ is defined by Eq.(\ref{W_from_Psi}).

\subsection{The ground state of the SUSY partner Hamiltonian $ \mathcal{H}_{+} $}
\label{sect_the_ground_state_H_plus} 

The first task is therefore to compute (approximately) the ground state WF  $ \Psi_0^{+}  $ of the SUSY partner Hamiltonian $ \mathcal{H}_{+} $.
To this end, note that a 
 candidate ground-state solution of $ \mathcal{H}_{+} $ 
is easy to compute from the equation $ \mathcal{A}^{+} 
\Psi_0^{+} = 0 $, whose solution can be obtained by flipping the 
sign of $ V(y) $ in Eq.(\ref{Psi_from_W}):
\beq
\Psi_0^{+}(y) \sim \frac{1}{\Psi_0(y)} \sim  \exp \left[   \frac{V(y) }{h^2} \right].
\eeq
However, this cannot be a right zero energy solution because it diverges as $ y \rightarrow \pm \infty $, and thus is not normalizable. 
The divergence can be removed if we consider the following ansatz \cite{SUSY_QM_Cooper, Junker}:
\beq
\label{SUSY_ansatz}
\Psi_{+} (y) = 
\left\{ \begin{array}{cc}
  \frac{1}{2 I_{+}} 
   \frac{1}{\Psi_0 (y)} \int_{y}^{\infty} dz 
\left[ \Psi_0 (z) \right]^2 & \text{for} \; y > 0   \\
     \frac{1}{2 I_{-}} 
    \frac{1}{\Psi_0(y)} \int_{-\infty}^{y} dz 
\left[ \Psi_0 (z) \right]^2 & \text{for} \; y < 0  \\
\end{array} \right.
\eeq
where
\beq 
I_{+} = \int_{0}^{\infty} dz \Psi_0^2(z) 
\; \; \;
I_{-} = \int_{-\infty}^{0} dz \Psi_0^{2}(z) 
\eeq  
The ansatz (\ref{SUSY_ansatz}) can now be computed in closed form using the ground state WF (\ref{Psi_0}).
Using Eq.(\ref{GM_3_comp}), we obtain
\beq
\label{SUSY_ansatz_GM}
\Psi_{+} (y) = 
\left\{ \begin{array}{cc} 
   \frac{1}{2 I_{+}} \frac{1}{ \Psi_0 (y)}  \sum_{k=1}^{3} \omega_k \mathcal{N} \left( \frac{\sqrt{2}(\mu_k T - y)}{ \sigma_k \sqrt{T} } \right), \; \; \; & \text{for} \; y > 0   \\
    \frac{1}{2 I_{-}} \frac{1}{\Psi_0(y)} 
 \sum_{k=1}^{3} \omega_k \mathcal{N} \left( \frac{ \sqrt{2}(y - \mu_k T) }{ \sigma_k \sqrt{T}} \right), \; \; \;      
     & \text{for} \; y < 0  \\
\end{array} \right.
\eeq
with 
\beq
\label{I_plus_minus_GM}
I_{+} =  \sum_{k=1}^{3} \omega_k \mathcal{N} \left( 
   \frac{\sqrt{2} \mu_k T }{ \sigma_k \sqrt{T}} \right), \; \; \; \;  I_{-} =  \sum_{k=1}^{3} \omega_k \mathcal{N} \left( - \frac{\sqrt{2} \mu_k  T}{  \sigma_k \sqrt{T}} \right) 
\eeq
These relations imply that $ \Psi_{+} (0) = \frac{1}{2 \Psi_0(0)} $. Note as as it stands, Eq.(\ref{SUSY_ansatz_GM}) is not very convenient for numerical implementation, as it can produce numerical overflows. A more convenient 
representation is obtained by using the representation of $ \Psi_{0} $ given in Eq.(\ref{Psi_0_new}), and expressing (\ref{SUSY_ansatz_GM}) in terms of the complimentary error function $  \text{erfc}(x) = 2 \mathcal{N}(- \sqrt{2} x ) $ and    
the scaled complementary error function $ \text{erfcx}(x) = e^{x^2} \text{erfc}(x) $:
\beq
\label{SUSY_ansatz_GM_2}
\Psi_{+} (y) = 
\left\{ \begin{array}{cc} 
  \frac{\sqrt{ 2 \pi \sigma_{+}^2}}{4 I_{+} C_{+} }    
 \sum_{k=1}^{3} \omega_k 
   e^{ \frac{(y-\mu_{+} T )^2}{2 \sigma_{+}^2 T } - \frac{(y-\mu_k T )^2}{\sigma_k^2 T }  - \log \eta(y)}
   \erfcx \left( \frac{y - \mu_k T }{ \sigma_k \sqrt{T} } \right), \; \; \; & \text{for} \; y > 0   \\
  \frac{\sqrt{ 2 \pi \sigma_{-}^2}}{4 I_{-} C_{-} }   \sum_{k=1}^{3} \omega_k 
   e^{ \frac{(y-\mu_{-} T )^2}{2 \sigma_{-}^2 T } - \frac{(y-\mu_k T )^2}{\sigma_k^2 T}  - \log \eta(y)}
   \erfcx \left(\frac{ \mu_k T - y}{ \sigma_k \sqrt{T}} \right), \; \; \;      
     & \text{for} \; y < 0  \\
\end{array} \right.
\eeq
where
\bea
\label{C_pm}
&&  C_{+} = C_{-} =  aC, \, \mu_{+} = \mu_{-} = \mu_2, \, \sigma_{+} = \sigma_{-} = \sigma_2 \; \;  \text{if} \; \; \sigma_2 > \sigma_1  \nonumber \\ 
&& C_{+} = C_{-} =  (1- a)C, \, \mu_{+} = \mu_{-}  = \mu_1, \, \sigma_{+}  = \sigma_{-} = \sigma_1 \; \;  \text{if} \; \; \sigma_2 < \sigma_1 \nonumber \\ 
&& C_{+}  = (1-a)C, C_{-} =  a C, \, \mu_{+} = \mu_1, \mu_{-} = \mu_2, \, \sigma_{+} = \sigma_1, \sigma_{-} = \sigma_2, \; \;  \text{if} \; \; \sigma_2 = \sigma_1, 
 \nonumber 
\eea 
As for $ x \rightarrow \infty$ $ \erfcx(x) \rightarrow (1/ \sqrt{\pi}) 1/x $ while $ \eta(y) \rightarrow 1$, this produces the following asymptotic behavior at $ y \rightarrow \pm \infty $ (assuming that $ \sigma_2 > \sigma_1 $):
\beq
\label{asymptotics_Psi_plus}
\Psi_{+} (y) = 
\left\{ \begin{array}{cc} 
   \frac{a}{4 \pi^{1/4} I_{+} \sqrt{ \Omega}}  \frac{1}{y - \mu_2 T } e^{ - (y-\mu_2 T )^2 /(2 \sigma_2^2 T )} , \; \; \; & \text{for} \; y  \rightarrow \infty   \\
   \frac{1-a}{4 \pi^{1/4} I_{-} \sqrt{ \Omega}}  \frac{1}{\mu_2 T - y} e^{ - (y-\mu_2 T)^2/(2 \sigma_2^2 T)} , \; \; \;      
     & \text{for} y \; \rightarrow - \infty  \\
\end{array} \right.
\eeq
For numerical integration of $ \Psi_{+} (y) $ that would be needed below, multiple calls for function $  \text{erfcx}(x)  $ can be time-consuming. As an alternative, one can rely on approximate expressions for $  \text{erfcx}(x)  $, see e.g. \cite{Ren_2007}. Other ways to simplify calculations will be discussed below.

Note that 
Eqs.(\ref{asymptotics_Psi_plus}) imply that $ \Psi_{+} $ is square-integrable, as it decays as a Gaussian multiplied by $ 1/y $.
The last relation is actually independent of a particular choice of the potential $ V(y) $, as the same functional form of a Gaussian times $ 1/y $ can also be found if the integrals 
in (\ref{SUSY_ansatz}) are evaluated using a saddle-point approximation around a minimum of a general potential $ V(y) $.
Assuming that a potential $ V(y) $ has a local minimum at point $ y_{\star} $ with a value $ V_{\star} $ and a second derivative $ V_{\star}'' $ at this point, 
a saddle-point calculation of integrals entering Eq.(\ref{SUSY_ansatz}) produces the following result for $ y > 0 $ (here $ u = \sqrt{V_{\star}''}(y - y_{\star})/h $):
\beq
\label{Psi_plus_saddle_point}
\Psi_{+} (y) =  \frac{1}{Z} e^{\frac{1}{2} u^2} \mathcal{N} \left( - \sqrt{2} u \right) 
 =    \frac{1}{2 Z}e^{- \frac{1}{2} u^2}  \erfcx  \left( u \right) 
 \underset{ y \rightarrow  \infty}{\longrightarrow} \frac{1}{Z} \frac{1}{y - y_{\star}}  e^{ - \frac{1}{2} \frac{(y - y_{\star})^2}{h^2/ V_{\star}''}} 
 \eeq
 where $ Z $ is a normalization constant whose specific expression will not be needed below, 
 and   $ \text{erfcx}(x) = e^{x^2} \text{erfc}(x) $  is the scaled complementary error function. For negative values $ y < 0 $, one obtains the same expression with 
 a flipped sign of the argument of $ \mathcal{N}(x), \,  \erfc(x), \,  \erfcx(x)  $, with $ y_{\star} $ substituted by a local minimum for the region $ y < 0 $. 
 Assuming that local minima of $ V(y) $ are well-separated for 
 the ground state WF (\ref{Psi_0}), their positions are well approximated by values $ \mu_1 $ and $ \mu_2 $ for $ x < 0 $ and $ x > 0 $, respectively, with $ V(\mu_1)'' = h^2/ (\sigma_1^2 T) $ and $  V(\mu_2)'' = h^2/ ( \sigma_2^2 T) $, respectively. 
  
Most of the calculation in this section do not however use the explicit form (\ref{SUSY_ansatz_GM}) and proceed with a general definition in 
Eq.(\ref{SUSY_ansatz}), while only relying on the asymptotic behavior (\ref{asymptotics_Psi_plus}).

One can easily check that $ \Psi_{+}(y) $ defined in Eq.(\ref{SUSY_ansatz}) is continuous at $ y = 0 $
with $ \Psi_{+} (0) = \frac{1}{2 \Psi_0(0)} $, and that
$ \mathcal{H}_{+} \Psi_{+}  = 0 $ for $ y \neq 0 $. For example, by taking $ y > 0 $, we obtain
\bea
\label{H+Psi}
\mathcal{H}_{+} \Psi_{+}(y) &=& \mathcal{A} \left[ - \frac{h^2}{\sqrt{2}} \frac{d \Psi_{+} (y) }{dy}  + \mathcal{W} \Psi_{+} (y) \right] = 
 \mathcal{A} \left[- \mathcal{W} \Psi_{+} (y)  + \frac{h^2}{\sqrt{2}} \frac{1}{2 I_{+} } \Psi_{0} (y)   + \mathcal{W} \Psi_{+} (y) \right] 
 \nonumber \\
 &=&   \frac{h^2}{\sqrt{2}} \frac{1}{2 I_{+} } \mathcal{A}  \Psi_{0} (y) = 0
 \eea
However, its derivative has 
a discontinuity at $ y = 0 $:
\beq
\label{discontinuity}
\left. \lim_{\varepsilon \rightarrow 0} \frac{d\Psi_{+} (y)}{dy} \right|_{\varepsilon} 
- \left. \frac{d \Psi_{+} (y)}{dy} \right|_{ - \varepsilon} = 
\lim_{\varepsilon \rightarrow 0} 2 \varepsilon 
\left. \frac{d^2 \Psi_{+} (y)}{dy^2} \right|_{y=0} =
- \frac{ \Psi_0(0) }{2} \left[ \frac{1}{I_{+}} + \frac{1}{I_{-}}
\right] = - \frac{ \Psi_0(0) }{2 I_{+} I_{-}}
\eeq 
Therefore, $\Psi_{+} $ defined in Eq.(\ref{SUSY_ansatz}) is not an eigenstate of $ \mathcal{H}_{+} $. 
Instead, $\Psi_{+}  $ is a zero-energy ground state WF
of a singular Hamiltonian $ \mathcal{H}_{s} $ given by
\beq 
\label{singular_H}
\mathcal{H}_{s} = \mathcal{H}_{+} - 
\frac{2 h^4}{  I_{+} I_{-}}
 \Psi_0^{2}(0)  \,  \delta(y)
\equiv \mathcal{H}_{+} - \delta  \mathcal{H},
\eeq 
that differs from  $ \mathcal{H}_{+} $ by the $ \delta $-function
term $ \delta  \mathcal{H} $. Indeed, with this Hamiltonian we obtain 
\beq 
\label{Normalizaton_SUSY_H_p}
\int dy  \Psi_{+} (y)\mathcal{H}_{s}
\Psi_{+} (y) = 0, 
 \eeq 
as a result of the fact that $ \mathcal{H}_{s} \Psi_{+} (y) =  \mathcal{H}_{+} \Psi_{+}  = 0 $ for 
$ y \neq 0 $, while finite contributions coming from the discontinuity of the first derivative (\ref{discontinuity}) and 
the additional $ \delta $-function
term cancel each other:
\beq 
\label{cancellation}
- \frac{h^4}{2}\int_{-\varepsilon}^{\varepsilon}
 \Psi_{+} (y)  \frac{d^2 \Psi_{+} (y)}{dy^2} dy - 
\int_{-\varepsilon}^{\varepsilon} \left[ \Psi_{+} (y) \right]^2 
\delta  \mathcal{H} dy  =
\frac{h^4}{8} \frac{1}{I_{+} I_{-}} -
\frac{h^4}{8}\frac{1}{I_{+} I_{-}}  = 0
\eeq 
Equivalently, we can express $ \mathcal{H}_{+} $ in terms 
of $ \mathcal{H}_{s} $:
\beq 
\label{H_singular_H}
\mathcal{H}_{+} 
= \mathcal{H}_{s} + \delta \mathcal{H}, \; \; \;  \delta \mathcal{H} :=  \alpha \delta(y), \; \; \; 
\alpha :=  \frac{2 h^4}{I_{+} I_{-}} \Psi_0^{2}(0) 
\eeq 

\subsection{Logarithmic Perturbation Theory for the log-wave function}
\label{sect_LPT_log_WF}

Given the structure of the Hamiltonian $ \mathcal{H}_{+} $ in Eq.(\ref{H_singular_H}), its ground-state eigenvalue and the eigenfunction 
can be found using quantum mechanical perturbation theory (see e.g. \cite{Landau_QM}).
To this end, we treat the additional term $ \delta \mathcal{H} $ as a perturbation  around 
the exactly solvable singular Hamiltonian $ \mathcal{H}_{s} $.
Following the literature on SUSY QM (see e.g. \cite{KKS, SUSY_tunneling_asymmetric}), 
we use the Logarithmic Perturbation Theory (LPT), instead of a more conventional  Rayleigh-Schr{\"o}dinger (RS) perturbation theory.
 For a one-dimensional Schr{\"o}dinger 
equation, the LPT yields quadrature formulas for energies and wave functions 
to any given order in the perturbation theory, in a hierarchical scheme.
The LPT becomes especially simple for a ground state which
is nodeless, and thus can be represented in the form $ \psi(x) = 
\exp [-G(x)/ \hbar ] $ where $ G(x) $ is some continuous function. By inversion of this formula, 
we have $ G(x) = - \hbar \log \psi(x) $, therefore function $ G(x) $ can be referred to as a log-wave function (log-WF). The 
perturbation 
scheme is then constructed for $ g(x) = G'(x) =  - \hbar \psi'(x) / \psi(x) $. 
While the LPT can be shown to be equivalent to the RS perturbation 
theory \cite{LPT}, it is far more convenient in practice due to its recursive quadrature form and the absence of a need  
to compute matrix elements of a perturbation potential. Details of the LPT method are provided in  Appendix C.

To the first order in perturbation $ \delta \mathcal{H} $, the ground-state energy $ E_0^{+} = E_1^{-} $ is 
given by Eq.(\ref{energy1order}) where we have to substitute $ \hbar \rightarrow h^2, \,  \alpha \rightarrow  2 h^4 \Psi_0^2(0)/(I_{+} I_{-}) $, 
and $ V_1(y) = \delta(y) $.
This gives the same result as the conventional RS perturbation theory (see e.g. \cite{Landau_QM}):
\bea   
\label{E_0_pt}
&& E_0^{+} = E_1^{-} = \alpha \bar{E}_1, \; \; \; \;  \alpha =   \frac{2 h^4 \Psi_0^2(0)}{I_{+} I_{-}}\\  
&& \bar{E}_1 = 
\frac{\int_{-\infty}^{\infty} 
 \Psi_{+}^2(y) \delta(y)  dy }{ 
\int_{-\infty}^{\infty} 
 \Psi_{+}^2(y) dy 
 }  
 = \frac{\Psi_{+}^2(0)}{\int_{-\infty}^{\infty} 
 \Psi_{+}^2(y) dy }
 \simeq
 \frac{\Psi_{+}^2(0)}{\int_{-\infty}^{\infty} e^{- \frac{2 V}{h^2}} dy  
 \int_{-\infty}^{\infty} e^{ \frac{2 V}{h^2}}dy } \nonumber 
 \eea 
Here the last approximate equality is obtained by assuming that the barrier is high, and therefore the ground state wave function $ \Psi_0 $ is concentrated around the minimum of $ V(y) $ where the exact form of $ V(y) $ is replaced by its approximation around this minimum. 
In other words, tunneling is neglected when we compute the energy splitting formula 
(\ref{E_0_pt}), similarly to how tunneling is computed in quantum mechanics (\cite{Landau_QM}, Sect. 50). With such assumption,
 $ \Psi_{+}(y) $ can be approximated as follows:
 $ \Psi_{+}(y) \simeq 1/ \Psi_0^{-}(y) = 1/C \exp\left[ 
 \frac{V(y)}{h^2} \right] $, where $ V(y) $ can be approximated by its quadratic expansion around it maximum. The first integral in the denominator 
 in the last expression in (\ref{E_0_pt}) comes due to the square of the normalization factor $ C $. 
 As the inverse of expression given in (\ref{E_0_pt}) is proportional to $ \mathcal{T} $ defined in Eq.(\ref{Tau}), such a Gaussian approximation produces the 
 same relation (\ref{Kramers_rate_main}) for the escape rate $ \lambda = E_1^{-}/h^2 $.  
Therefore, the classical Kramers escape rate relation can also be obtained using methods of SUSY \cite{Brown, Junker}. 

For the ground-state wave function $ \Psi_{0}^{+}(y) $ of the Hamiltonian $ \mathcal{H}_{+} $, the LPT produces a perturbative expansion 
of a 'log-WF' $ G(y) = 
- h^2 \log \Psi_{0}^{+}(y) $ in powers of perturbation parameter $ \alpha =  2 h^4  \Psi_0^2(0) /(I_{+} I_{-}) $.
To the first order in $ \alpha $, the WF  $ \Psi_{0}^{+}(y)  $ has the following form:
\beq
\label{Psi_plus_from_G_1}
\Psi_{0}^{+}(y) = C_{1} e^{ -  \frac{G_0(y)}{h^2} -  \frac{\alpha G_1(y)}{h^2}} =  C_{1} \Psi_{+}(y) e^{ - \frac{\alpha G_1(y)}{h^2}}, \; \;  \alpha =  \frac{2 h^4}{I_{+} I_{-}} \Psi_0^{2}(0), 
\eeq 
where $ C_1 $ is a normalization factor to be determined later, $ \Psi_{+}(y) $ is the unperturbed WF defined in (\ref{SUSY_ansatz}), and $ G_0(y) = - h^2 \log \Psi_{+}(y)/ C_1  $ stands for the log-WF of the unperturbed Hamiltonian $ \mathcal{H}_s $. 

While a specific expression for the first-order correction $ G_1(y) $ will be given momentarily, one should note that within perturbation theory, a first-order approximation of the form (\ref{Psi_plus_from_G_1}) is only justified when a first correction $ \alpha G_1(x) $ is smaller than  
an unperturbed log-WF $ G_{0}(x) $:
\beq
\label{applicability_of_pert_th}
 \alpha  \left| G_1(x) \right| <  \left| G_{0}(x) \right|, \; \; \; \;  G_{0}(x) = - h^2 \log \frac{\Psi_{+}(x)}{ C_1} 
 \eeq
As implied by Eq.(\ref{asymptotics_Psi_plus}), function $ G_{0}(x)  $ grows quadratically in the asymptotics: $ G_{0}(x) \propto x^2 $ when $ x \rightarrow \pm \infty $.  On the other hand, as the perturbation potential 
in our problem is given by a Dirac delta-function centered at $ x = 0 $, the function $ G_1(x) $ that incorporates its effect is expected to decay as 
$  x \rightarrow \pm \infty $. Therefore, using perturbation theory is justified in our problem.

\begin{figure}[ht]
\begin{center}
\includegraphics[
width=160mm,
height=50mm]{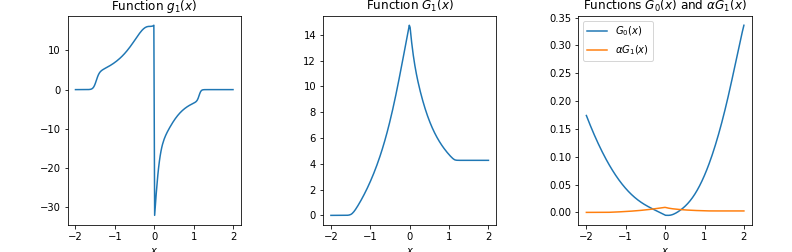}
\caption{Functions $ g_1(x) = G_1'(x) $ (the left graph) and $ G_1(x) $ (the graph in the center) obtained by numerical integration of $ \Psi_{+}(y) $ according to Eqs.(\ref{g_1_new}) and 
(\ref{G_1_new}). The following parameters are used $ a = 0.3,  \, \mu_1 = 0.4, \, \mu_2 = -0.4, \, \sigma_1 = 0.2, \, \sigma_2 = 0.3, \, 
h = 0.05 $. The graph on the right shows the scaled function $ \alpha G_1(x) $ vs the log-WF $ G_{0}(x) = - h^2 \log \Psi_{+}(x) $ of the unperturbed Hamiltonian 
$ \mathcal{H}_{s} $, see Eq.(\ref{Psi_plus_from_G_1}). While $ G_1(x) $ decays as $ x \rightarrow \pm \infty $, $ G_{0}(x) $ grows as a quadratic function.}
\label{fig_g1_G1}
\end{center}
\end{figure}

 
The first-order correction log-WF  $ G_1(x) $ is determined by the following relations,  see Eqs.(\ref{ansatz_LPT}) and (\ref{WF1order}):
\beq
\label{G_1_fun}
G_1 (x) = 
\left\{ \begin{array}{cc} 
 \frac{2}{ h^2}  \int_{-\infty}^{x}  \frac{dy }{\Psi_{+}^{2} (y) } \int_{-\infty}^{y} dz \, 
\left( \bar{E}_1  - \delta (z)  \right) \Psi_{+}^{2}(z) , \; \; \; &  x > 0   \\
- \frac{2}{ h^2}   \int_{-\infty}^{x}   \frac{dy}{\Psi_{+}^{2} (y) } \int_{y}^{\infty} dz \, 
\left( \bar{E}_1  - \delta (z) \right) \Psi_{+}^{2}(z)  ,  \; \; \;   &  x < 0  \\
\end{array} \right. 
\eeq
We omit here an additive constant that can re-absorbed into the normalization constant $ C_1 $ in 
(\ref{Psi_plus_from_G_1}). 
Using the constraint (\ref{E_0_pt}), Eq.(\ref{G_1_fun}) can be written in a different form:
\beq
\label{G_1_new}
G_1 (x) =   
\left\{ \begin{array}{ll} 
   \frac{2 \bar{E}_1}{h^2}  \left[ \int_{-\infty}^{0} \frac{ dy}{\Psi_{+}^2(y)} \int_{-\infty}^{y} dz \Psi_{+}^2(z) - 
 \int_{0}^{x} \frac{ dy}{\Psi_{+}^2(y)} \int_{y}^{\infty} dz \Psi_{+}^2(z)  \right] 
   , \; \; \; & \text{for} \; x > 0   \\
   \frac{2 \bar{E}_1}{h^2}   \int_{-\infty}^{x} \frac{ dy}{\Psi_{+}^2(y)} \int_{-\infty}^{y} dz \Psi_{+}^2(z)
 , \; \; \;      
     & \text{for} \; x < 0  \\
\end{array} \right.
\eeq
Due to Eq.(\ref{E_0_pt}), $ G_1(x) $ is continuous at $ x = 0 $ in Eq.(\ref{G_1_fun}), as can also be seen in Eq.(\ref{G_1_new}). 
For the derivative $ g_1(x) = G_1'(x) $, we find 
\beq
\label{g_1_new}
G_1' (x) =   
\left\{ \begin{array}{ll} 
   - \frac{2 \bar{E}_1}{h^2}   \frac{1}{\Psi_{+}^2(x)}   \int_{x}^{\infty} dz \Psi_{+}^2(z)  
   , \; \; \; & \text{for} \; x > 0   \\
     \frac{2 \bar{E}_1}{h^2}   \frac{1}{\Psi_{+}^2(x)}   \int_{-\infty}^{x} dz \Psi_{+}^2(z)  
 , \; \; \;      
     & \text{for} \; x < 0  \\
\end{array} \right.
\eeq
 The derivative $ G_1'(x) $ is positive for $ x < 0 $ and negative for $ x > 0 $, see Fig.~\ref{fig_g1_G1}.  Therefore
$ G_1(x) $ has a maximum at $ x = 0 $ with the value
\beq
\label{G_1_max}
G_1(0) =  \frac{2  \bar{E}_1}{ h^2}   \int_{-\infty}^{0}  \frac{dy }{\Psi_{+}^{2} (y) } \int_{-\infty}^{y} dz \,  \Psi_{+}^{2}(z)  >  0
\eeq 
Note that while $ G_1(x) $ is continuous at $ x = 0 $, its second derivative has a jump: for a small value  $ \varepsilon 
\rightarrow 0 $, we obtain 
\beq
\label{G_pp_jump}
G_1''(\varepsilon) - G_1''(-\varepsilon) = \frac{2 \bar{E}_1}{h^2} + \frac{2 \bar{E}_1}{h^2}  = \frac{4 \bar{E}_1}{h^2}, \; \; \;  \varepsilon 
\rightarrow 0 
\eeq

\subsection{The first excited state of $ \mathcal{H}_{-} $}
\label{sect_the_first_excited_state} 

When the ground state WF  of the SUSY partner Hamiltonian $ \mathcal{H}_{+} $ is computed to the first order of LPT in Eq.(\ref{Psi_plus_from_G_1}), we can now use SUSY relations (\ref{SUSY_relations_main}) to obtain excited states of $ \mathcal{H}_{-} $. For the first excited state 
$ \Psi_1^{-} $ of $ \mathcal{H}_{-} $, we have the following relation in terms of the ground state $ \Psi_0^{+} $ of $ \mathcal{H}_{+} $:
\beq
\label{SUSY_relation_Psi_1}
 \Psi_{1}^{-} =  \left(E_0^{+} \right)^{-1/2}
\mathcal{A}^{+}\Psi_{0}^{+}  =  \left(E_0^{+} \right)^{-1/2} \left[  - \frac{h^2}{\sqrt{2}} \frac{d \Psi_{0}^{+}(y)  }{dy}  + \mathcal{W} \Psi_{0}^{+}(y) \right] 
\eeq
Substituting here Eqs.(\ref{SUSY_relations_main}) and.(\ref{Psi_plus_from_G_1}), we obtain
\beq
\label{First_excited}
 \Psi_{1}^{-}(y)  = 
\left\{ \begin{array}{cc} 
\frac{\left(E_0^{+} \right)^{-1/2} }{\sqrt{2}} \frac{C_1 h^2 }{2 I_{+}} 
 \left[ \Psi_0(y) e^{ - \frac{\alpha G_1(y)}{h^2}} - \frac{1}{\Psi_0(y)} 
   \frac{d}{dy} \left(   e^{ - \frac{\alpha G_1(y)}{h^2}} \right) \int_{y}^{\infty} dz \Psi_0^2(z) \right]
   , \; \; \; &  y > 0   \\
\frac{\left(E_0^{+} \right)^{-1/2} }{\sqrt{2}} \frac{C_1 h^2 }{2 I_{-}} 
   \left[- \Psi_0(y) e^{ - \frac{\alpha G_1(y)}{h^2}} - \frac{1}{\Psi_0(y)} 
   \frac{d}{dy} \left(   e^{ - \frac{\alpha G_1(y)}{h^2}} \right) \int_{-\infty}^{y} dz \Psi_0^2(z) \right]
   ,  \; \; \;   &  y < 0  \\
\end{array} \right.   
\eeq
We can check that this WF is orthogonal to the ground state WF $ \Psi_0 = \Psi_0^{-} $:
\beq
\label{orthogonality}
\int_{-\infty}^{0} dy \Psi_0^{-} (y) \Psi_{1}^{-} (y)  + 
\int_{0}^{\infty} dy \Psi_0^{-} (y) \Psi_{1}^{-} (y) 
= - \frac{C_1 h^2}{2 \sqrt{2}} e^{ - \frac{\alpha G_1(0)}{h^2}} +  \frac{C_1 h^2}{2 \sqrt{2}} e^{ - \frac{\alpha G_1(0)}{h^2}} = 0
\eeq
One can also check that the first state $ \Psi_1^{-} $ is squared-normalized provided $ \Psi_{0}^{+} $ is square-normalized, while the coefficient $  \left(E_0^{+} \right)^{-1/2} $ takes care of a proper normalization of $ \Psi_1^{-} $. 
 
\begin{figure}[ht]
\begin{center}
\includegraphics[
width=170mm,
height=110mm]{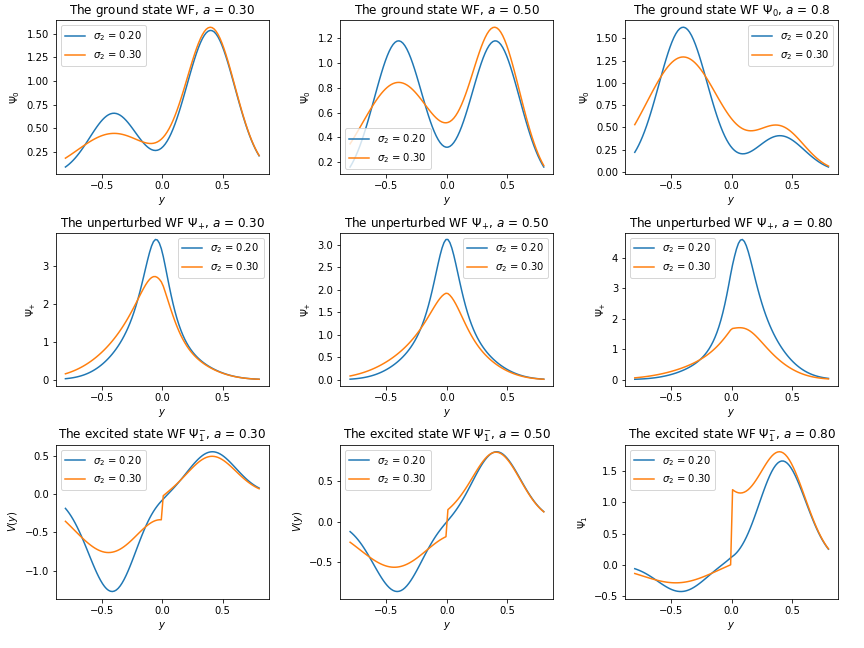}
\caption{The ground state WF $ \Psi_{0} $ of the Hamiltonian $ \mathcal{H}_{-} $, the unperturbed ground state $ \Psi_{+} $ of the SUSY partner Hamiltonian 
$ \mathcal{H}_{+} $, and the first excited state WF $ \Psi_{1}^{-} $ of   $ \mathcal{H}_{-} $. } 
\label{fig_first_second_WFs}
\end{center}
\end{figure}

Shapes of $ \Psi_1^{-} $ obtained in our model are illustrated in Fig.~\ref{fig_first_second_WFs}, along with a ground state WF $ \Psi_{0} =  \Psi_{0}^{-} $, and 
an unperturbed ground state WF $ \Psi_{+} $ of the partner SUSY Hamiltonian $ \mathcal{H}_{+} $. As could be expected, the WF $ \Psi_1^{-} $ of the first excited state resembles the well known WF $ \Psi_1 \propto x e^{ - m \omega x^2/2 \hbar} $ of the first excited state of the  quantum mechanical harmonic oscillator \cite{Landau_QM}.

Once the first state  $ \Psi_1^{-} $ is computed using SUSY, we are done in terms of computing the leading pre-asymptotic correction to a stationary distribution.
It is given by Eq.(\ref{two_terms_QM}) which we repeat here for convenience:
\beq
\label{two_terms_QM_2}
p(y,t | y_0) =    \Psi_0^{2} (y) + 
 \frac{ \Psi_1^{-}(y_0)}{ \Psi_0 (y_0)} e^{- \frac{ t  \Delta E}{h^2}}  \Psi_0 (y) \Psi_1^{-}(y) + \ldots  
\eeq
This means that the non-stationary component of the return distribution is given, at a leading pre-asymptotic order, by the product  $  \Psi_1^{-}(y) \Psi_1^{-}(y_0) $.
This immediately translates into non-stationary corrections to the variance and skewness of the return distribution, as will be discussed in the next section. 

\section{Practical implementation and applications to option pricing}
\label{sect_Applications}

\subsection{NES: Practical implementation}
\label{sect_practical_version}

Formulas presented above provide an accurate theoretical scheme of computing the first excited state WF $  \Psi_1^{-} $ in Eq.(\ref{First_excited}), and hence
give an explicit solution for a time-dependent component in the non-stationary return distribution  (\ref{two_terms_QM_2}). 
In practice, computing this WF involves a numerical computation of a double integral that defines function $ G_1(x) $, see Eq.(\ref{G_1_fun}), which may be time-consuming for model calibration when such computations should be done many times. One way to handle such potential challenges that was mentioned above was to use fast numerical approximations to evaluate 
the scaled complimentary error function $ \erfcx(x) $, however this still requires a double numerical integration.

An alternative that might be preferred in practice would be to directly approximate the WF $  \Psi_1^{-} $. As was mentioned above, it resembles, not accidentally, the first excited state WF $ \Psi_1 \propto x e^{ - m \omega x^2/2 \hbar} $ of a quantum mechanical harmonic oscillator.  This suggests that a simple 
approximation  $  \Psi_1^{-} $ would be given by the following expression:
\beq
\label{Psi_1_simple}
 \Psi_1^{-}(y) \simeq \frac{1}{\sigma_M} (y - \bar{y}_M ) \Psi_{0}(y) 
 \eeq
 where $ \bar{y}_M $  and $ \sigma_M $ are the mean and standard deviation of the Gaussian mixture defined by $ \Psi_{0}^2 $:
 \beq
 \label{Z_Psi_1}
  \bar{y}_M = \int dy y \Psi_{0}^2(y) = \sum_{k=1}^{3} \omega_k \mu_k  T, \; \; \; 
 \sigma_M^2 =   \int dy (y- \bar{y}_M)^2 \Psi_{0}^2(y)  = \sum_{k=1}^{3} \omega_k \left( \mu_k^2 T^2 + \frac{\sigma_k^2}{2} T \right) -  \bar{y}_M^2
 \eeq
 Substituting (\ref{Psi_1_simple}) into Eq.(\ref{two_terms_QM_2}) produces the following simplified form that can be used for practical applications of the model:
 \beq
\label{two_terms_QM_3}
p(y,t | y_0) =   \Psi_0^2 (y) \left[ 1  + 
 b_t (y - \bar{y}_M) \right], \; \; \; b_t :=   \frac{ y_0 -  \bar{y}_M }{ \sigma_{M}^2 } e^{-  \lambda  t } 
\eeq
where we used the definition $ \lambda = \frac{ \Delta E}{h^2} $ of the Kramers escape rate, where the energy splitting $  \Delta E $ is computed according to Eq.(\ref{E_0_pt}).  Note that the density defined by this relation can formally become negative for large negative values of $ y - \bar{y}_M $. However, this is an artifact of a reliance on the first-order perturbation theory that was used to derive Eq.(\ref{two_terms_QM_3}). As long as perturbation theory is applicable, 
we  have $ b_t(y - \bar{y}_M) \ll 1 $, and can thus replace $ 1 + b_t(y -  \bar{y}_M) \rightarrow e^{b_t(y -  \bar{y}_M)} $ under the same assumptions.  Therefore, we can replace (\ref{two_terms_QM_3}) by even a simpler form
\beq
\label{two_terms_QM_4}
p(y,t | y_0) =  C  e^{b_t y }  \Psi_0^2 (y) 
\eeq
where the factor $ e^{ - b_t \bar{y}_M} $ is absorbed into a normalization constant $ C $. When perturbation theory applies, differences between Eqs.(\ref{two_terms_QM_4}) and (\ref{two_terms_QM_3}) would be negligible as long as  all integrals 
defining observable quantities using the definition (\ref{two_terms_QM_4}) are dominated by a region of small $  y $ where $ b_t | y - \bar{y}_M |  \leq 1 $. 
In this case, using (\ref{two_terms_QM_4}) instead of (\ref{two_terms_QM_3}) would simplify calculations, while any differences could be re-absorbed in slight modifications of model parameters relative to their values in the original equation (\ref{two_terms_QM_3}). 

While Eq.(\ref{two_terms_QM_4}) will be used below for option pricing, it is more illuminating to use the original form (\ref{two_terms_QM_3}) to 
approximate time-dependent contributions to moments of the return distribution. For the mean, variance and skewness of the time-dependent distribution (\ref{two_terms_QM_3}), we obtain
\bea
\label{mean_var_skew_time_dependent}
&& \mathbb{E} \left[ y_t \right] =  \bar{y} +  b_t \sigma_M^2,  \; \; \; \sigma_M^2 =   \int dy (y- \bar{y})^2 \Psi_{0}^2(y)  \nonumber \\
&&  \mathbb{E} \left[ \left(y_t - \bar{y} \right)^2 \right] 
=  \sigma_M^2  +  b_t  \mathcal{M}_3, \; \; \;  \mathcal{M}_3 = \int dy ( y - \bar{y})^3 \Psi_{0}^2(y) \\
&&  \mathbb{E} \left[  \left(y_t - \bar{y} \right)^3 \right] =   \mathcal{M}_3 +    b_t  \mathcal{M}_4, \; \; \;  \mathcal{M}_4 = \int dy ( y - \bar{y})^4 \Psi_{0}^2(y)  \nonumber 
\eea
Here $  \mathcal{M}_3 $ and $  \mathcal{M}_4 $ are centered moments computed from the three-component Gaussian mixture 
$ \Psi_{0}^2 $ according to formulas in Appendix D:
\beq
\label{uncentered_moments_main}
 \int dy  y^3  \Psi_{0}^2(y) = \sum_{k=1}^{3} \omega_k \left(\mu_k^3 T^3 + \frac{3}{2} \mu_k \sigma_k^2 T^2 \right), \; \; \; 
 \int dy  y^4  \Psi_{0}^2(y)= \sum_{k=1}^{3} \omega_k \left(\mu_k^4 T^4 + 3 \mu_k^2 \sigma_k^2 T^3 + \frac{3}{4}  \sigma_k^4 T^2\right) 
\eeq
Equations.(\ref{mean_var_skew_time_dependent}) are applicable as long as corrections $ \sim b_t $ are smaller than the first terms in these formulae. Interestingly, the model predicts that a time-dependent correction to the $ n$-th moment of a stationary distribution is approximately proportional to the $(n+1)$-th moment, where the coefficient of proportionality has the same time dependence for all moments.  This observation could shed some light on some open questions 
in financial modeling. In particular, \cite{Bouchaud_Skew_risk_premia} suggests that risk premia in equity returns are in fact premia for the skewness, rather than variance of returns. This goes contrary to the traditional financial theory such as CAPM \cite{CAPM} that claims that risk premia are premia for the variance. Our 
Eqs.(\ref{mean_var_skew_time_dependent}) show that the variance and skewness of a stationary distribution are mixed in a time-dependent way in a non-equilibrium, pre-asymptotic distribution, thus suggesting that both views may be partially correct, and be in fact parts of the same time-dependent analysis. A more detailed look into implications for risk premia is left here for a future research. 

\subsection{Non-equilibrium option pricing: construction of the pricing measure}
\label{sect_option_pricing}

In addition to providing explicit formulae for time-dependent moments of returns, the analytical approach of this paper can also be applied to option pricing in a non-equilibrium, pre-asymptotic setting.
To this end, we use the simplified form (\ref{two_terms_QM_4}) of a finite horizon transition density.  To use it for option pricing, recall that $ y_t $ is defined as 
a time-$t$ log-return computed over the time period $ [t-T,t] $:
\beq
\label{y_t_S_t}
y_t = \log \frac{S_t}{S_{t-T}}
\eeq 
so that $ y_T = \log S_T/S_0 $. 
Using Eq.(\ref{two_terms_QM_4}), the transition density for $ y_T $ can be written as follows:
\beq
\label{trans_density_y_T}
p(y_T | y_0) =   C  e^{b y_T }  \Psi_0^2 (y)  = C   e^{b y_T}  \sum_{k=1}^{3} \omega_k \phi \left(y_T | \mu_k T,  \hat{\sigma}_k^2 T \right), \; \; \; \hat{\sigma}_k^2 = \frac{\sigma_k^2}{2}, \; \; 
b =   \frac{ y_0  -  \bar{y}_M}{ \sigma_{M}^2 } e^{-  \lambda  T } 
\eeq
where $ C $ is a normalization constant, and we write $ b = b_T $ to lighten the notation. Eq.(\ref{trans_density_y_T}) approximates pre-asymptotic effects in the time-$T$ distribution of log-return $ y_T $, where 
such effects are controlled by parameter $ b = b_T$. We can write it as another GM distribution with different weights and means:
\beq
\label{p_real_measure}
p(y_T) =   \sum_{k=1}^{3} \omega_k^{(p)} \phi \left(y_T | \mu_k^{(p)} T,  \hat{\sigma}_k^2 T \right)
\eeq
where we omitted the dependence on $ y_0 $ for brevity, and defined the adjusted parameters as follows:
\beq
\label{mu_p}
\mu_k^{(p)} := \mu_k +  b \hat{\sigma}_k^2, \; \; \; \omega_k^{(p)}  :=  \frac{ \hat{\omega}_k}{\sum_{k} \bar{\omega}_k}, \; \; \; \hat{\omega_k}  := 
\omega_k e^{ b \left( \mu_k +  b  \frac{\hat{\sigma}_k^2}{2} \right) T}
\eeq
The stationary distribution can be formally obtained from this expression by taking the limit  $ b_T \rightarrow 0 $.

The non-equilibrium density  (\ref{trans_density_y_T}) refers to a statistical (real-world) measure. Distributions needed for option pricing are risk-neutral distributions where by definition the expected return of a stock paying dividend $ q $ is given by $ r_f  - q $ where $ r_f $ stands for a risk-free rate. By It{\^o}'s lemma, the expectation value of log-price  $ y_T $ under the risk-neutral measure should then be equal to $ (r_f - q - \frac{{h}^2}{2}) T $.
A simple approach to construct a risk-neutralized distribution 
$ q (y_T) $ starting with a real-measure distribution $ p(y_T) $ is to apply the Minimum Cross Entropy method. With this method, $ q (y_T) $ is found by minimization of the following Lagrangian function:
\beq
\label{Lagrangian}
\mathcal{L} = \int d y_T q (y_T) \log \frac{ q (y_T)}{p(y_T)} - \xi \left( \int d y_T  y_T q (y_T)  - \left( r_f - q - \frac{{h}^2}{2} \right) T \right)
\eeq
where $ \xi $ is a Lagrange multiplier, and the second term is the constraint  ensuring that the expected value of $ y_T $ is equal to $  (r_f - q - \frac{{h}^2}{2}) T $.
This expression should be minimized with respect to $ q(y_T) $, and maximized with respect to $ \xi $. Notice that such a procedure would induce a dependence of the optimal value of $ \xi $ on the parameter $ h $.

 Minimization of this Lagrangian with respect to $ q(y_T) $ produces  the following expression for $ q(y_T) $:
\beq
\label{risk_neutral}
q(y_T) =  C' e^{ \xi y_T}  p(y_T) =  \frac{1}{Z_{\xi}}  e^{(\xi + b) y_T }  \Psi_0^2 (y_T) , \; \; \; 
Z_{\xi} := \int d y_T e^{ (\xi + b) y_T}  \Psi_0^2 (y_T) 
\eeq
where $ C' $ is another normalization factor that is re-absorbed into the new normalization factor  $  \frac{1}{Z_{\xi}} $ in the second equation. 
Such an exponential tilt of the real-world measure 
 $ p(y_T) $ to produce a risk-neutral measure $ q(y_T) $ is also known as the Esscher transform. Notice that $ q(y_T) $ depends on $ \xi $ only via the 
 combination $  \xi  + b $, which will be used below.  
 
 Plugging the solution (\ref{risk_neutral}) back into the Lagrangian 
 (\ref{Lagrangian}) and flipping the overall sign, the problem of finding the Lagrange multiplier amounts to minimization of the following objective function:
 \beq
 \label{Lagrangian_2}
\mathcal{L}' (\xi) = \log Z_{\xi}  - \xi \left( r_f - q - \frac{{h}^2}{2} \right) T = \log \sum_{k=1}^{3}  \omega_k e^{ (\xi + b) \left( \mu_k + 
( \xi + b) \frac{ \hat{\sigma}_k^2}{2} \right) T} 
- (\xi + b) \left( r_f - q - \frac{{h}^2}{2} \right) T + \text{const}
\eeq
with respect to $ \xi $. Equivalently but more conveniently, in the second equality in Eq.(\ref{Lagrangian_2}), we write the objective function as a function of 
a combination $ \xi' := \xi + b $, plus an unessential constant term equal to $  b \left( \frac{{h}^2}{2} -  r_f + q \right) T $). Given that both $ q(y_T) $ and 
the loss function (\ref{Lagrangian_2}) depend on $ \xi $ only via the combination $ \xi' = \xi + b $, we can equivalently, but more conveniently, replace optimization with respect to $ \xi $ in Eq.(\ref{Lagrangian_2}) by optimization with respect to $ \xi' $. As it is easy to see, setting the derivative of $ \mathcal{L}' (\xi')  $ to zero produces the constraint for the expected value of $ y_T $
\beq
\label{expected_return_constraint}
\int  y_T  q(y_T) dy_T  =  (r_f - q -  \frac{{h}^2}{2}) T 
 \eeq
Furthermore, the second derivative $  \partial^2 \mathcal{L}' (\xi) / \partial \xi^2  $ is non-negative as it is equal to the variance of $ y_T $, therefore the 
minimization problem in Eq.(\ref{Lagrangian_2}) is convex and has a unique solution for $ \xi' = \xi + b $.
 
The risk-neutral distribution $ q(y_T) $ in Eq.(\ref{risk_neutral}) can now be re-written as a three-component Gaussian mixture similar to the real-measure 
density (\ref{p_real_measure}), but with different means and weights:
\beq
\label{GM_3_comp_q}
 q(y_T)  =  \sum_{k=1}^{3} \omega_k^{(q)} \phi \left(y_T | \mu_k^{(q)} T, \hat{\sigma}_k^2 T \right) 
 \eeq
with 
\beq
\label{mu_rn}
 \mu_k^{(q)} = \mu_k +  (\xi + b)  \hat{\sigma}_k^2, \; \; \;  \omega_k^{(q)}  =  \frac{ \bar{\omega}_k}{\sum_{k} \bar{\omega}_k}, \; \; \; \bar{\omega_k}  = \omega_k e^{  (\xi + b) \left( \mu_k^{(p)} +  (\xi + b) \frac{\hat{\sigma}_k^2}{2} \right) T}
\eeq
so that $ \ \mu_k^{(p)} =  \left. \mu_k^{(q)} \right|_{\xi = 0} $ and  $ \omega_k^{(p)} = \left.  \omega_k^{(q)} \right|_{\xi = 0} $.
The constraint (\ref{expected_return_constraint}) can now be written as follows:
\beq
\label{expected_return_constraint_2}
 \sum_{k=1}^{3}  \omega_k^{(q)}  \mu_k^{(q)}  = r_f - q - \frac{{h}^2}{2}
 \eeq
Re-grouping terms, this equation can also be written as follows:
\beq
\label{expected_return_constraint_4}
\xi' = \frac{ \sum_{k=1}^3 \omega_k  \left( r_f- q - \frac{{h}^2}{2} - \mu_k \right) e^{ \xi' T \left( \mu_k + \xi' \frac{\hat{\sigma}_k^2}{2} \right)}}{
 \sum_{k=1}^3 \omega_k \hat{\sigma}_k^2 e^{ \xi' T \left( \mu_k + \xi' \frac{\hat{\sigma}_k^2}{2} \right)}}, \; \; \; \xi' := \xi + b_T
 \eeq
 This equation could be used to solve for $ \xi' $ by iterations, which could be considered an alternative to finding the optimal value of $ \xi' $ by 
 a direct numerical minimization of the Lagrangian (\ref{Lagrangian_2}).
 The limiting behavior at $ T \rightarrow 0 $ is
 \beq
 \xi'  \underset{ T \rightarrow  0}{\longrightarrow}   \frac{ \sum_{k=1}^3 \omega_k \left( r_f - q- \frac{{h}^2}{2} - \mu_k \right)}{   \sum_{k=1}^3 \omega_k    \hat{\sigma}_k^2}  
\eeq

Note again that the risk-neutral distribution $ q(y_T) $ in Eq.(\ref{GM_3_comp_q}) only depends on $ \xi $ via the combination $ \xi' =  \xi + b_T $.
On the other hand, the 
latter combination is fixed in terms of the original model parameters by Eq.(\ref{expected_return_constraint_4}). 
This fact implies, in particular, that while parameter $ b_T $ quantifies the amount of non-equilibrium in the system, for the purpose of finding the 
risk-neutral distribution $ q(y_T) $, its exact value is irrelevant as it gets re-absorbed into the only relevant parameter combination $ \xi' =  \xi + b_T $. 
It is only the real-measure distribution 
$ p(y_T) $ in Eq.(\ref{p_real_measure}) that explicitly depends on parameter $ b = b_T $, and hence is sensitive to the amount of non-equilibrium in the dynamics.

\subsection{European option pricing and model calibration}
\label{sect_option_pricing_2}

Now consider a European call option on $ S_T $ with maturity $ T $ and strike $ K $ and a terminal payoff $ (S_T - K)_{+} $. The option value is defined as a discounted expected value of its payoff function with respect to the risk-neutral measure (\ref{GM_3_comp_q}):
\beq
\label{option_price}
C(K,T) = e^{ - r_f T} \int_{-\infty}^{\infty}  \left(S_0 e^{y_T} - K \right)_{+}   q(y_T)  d y_T
\eeq
Plugging here Eq.(\ref{GM_3_comp_q}) and performing integration, we obtain a closed-form expression for the option price in a non-equilibrium setting:
\beq
\label{opt_price_res}
C(K,T) =\sum_{k=1}^{3} \omega_k^{(q)} \left[ S_0 e^{ - q_k T} \mathcal{N} \left(d_1^{(k)} \right) - 
K  e^{ - r_f T}   \mathcal{N} \left( d_2^{(k)} \right)  \right]
\eeq
where
\bea
\label{BS_mixture_params}
&& q_k = r_f - \mu_k - \left( \xi' + \frac{1}{2}\right) \hat{\sigma}_k^2, 
 \nonumber \\
&& d_1^{(k)}  = \frac{ \log \frac{S_0}{K} + \left(r_f - q_k +  \frac{1}{2} \hat{\sigma}_k^2 \right) T}{ \hat{\sigma}_k \sqrt{T} }, 
\; \;  d_2^{(k)} =  d_1^{(k)} -  \hat{\sigma}_k \sqrt{T} 
\eea
where parameter $ \xi' = \xi + b_T $ is found by the numerical minimization of Eq.(\ref{Lagrangian_2}), or alternatively by iterations in Eq.(\ref{expected_return_constraint_4}). 
Each component in the sum in Eq.(\ref{opt_price_res}) is given by the classical Black-Scholes formula \cite{BS} for the price of a dividend-paying stock with  dividend $ q_k $. To differentiate them from real dividends paid on stocks, in what follows we will refer to parameters $ q_k $ as the NES dividends. 

It should be noted here that mixtures of lognormal price distributions, or equivalently normal mixtures for returns were previously suggested in the literature on empirical grounds as a flexible way to incorporate volatility smiles in option prices, see e.g. \cite{Alexander}.\footnote{ Two-component Gaussian mixtures were also  proposed for portfolio optimization in \cite{Roncalli_2019} as a phenomenological way to incorporate skewness of equity returns.}  
The novelty of the approach presented in this paper is in providing a theoretical foundation to such mixture models, while simultaneously providing a parsimonious representation of a three-component Gaussian mixture in terms of only 5 model parameters $ \mu, \sigma_1, \sigma_2, a, h $.
Note that the option price depends on the value of parameter $ h $ only through its dependence on the NES dividends $ q_k $, which in their turn depend on $ \xi' $.
Furthermore, once parameters $ \theta = \left(  \mu, \sigma_1, \sigma_2, a, h \right) $ are found from calibration to option prices, they produce \emph{both} the risk-neutral and real-measure return distributions.  

The availability of the closed form solution (\ref{opt_price_res}) for European call options (and a similar relation for puts) enables a straightforward calibration of model parameters  $ \theta = (\mu, \sigma_1, \sigma_2, a, h ) $ to market option data.\footnote{In addition to option pricing data, one can use other types of data for model calibration, as will be discussed below.} To this end, one should optimize a loss function penalizing the difference between market option prices and their theoretical expressions given for calls options by Eq.(\ref{opt_price_res}) as functions of   $ \theta = ( \mu, \sigma_1, \sigma_2, a, h ) $, with an additional regularization as described in more details in the next section. As the effective dividend parameters 
$ q_k $ in Eq.(\ref{BS_mixture_params}) depends on the parameter combination  
 $ \xi' = \xi + b_T $ which is defined in terms of the original model parameters according to Eq.(\ref{expected_return_constraint_4}), a new value of parameter
  $ \xi' $ should be computed at each iteration of optimization of the model parameter $ \theta $.


The procedure just outlined allows one to calibrate the risk-neutral  distribution (\ref{GM_3_comp_q}) by fitting the model parameter $ \theta = (\mu, \sigma_1, \sigma_2, a, h ) $. As was discussed above, this distribution is also a function of  $ \xi' = \xi + b_T $ fixed in terms of parameters $ \theta $ by  
Eq.(\ref{expected_return_constraint_4}). The actual value of parameter $ b_T $ is irrelevant for the risk-neutral distribution  (\ref{GM_3_comp_q}) as it gets re-absorbed into the adjusted parameter $ \xi' = \xi + b_T $.

While the risk-neutral distribution $ q(y_T) $ is insensitive to variations of parameter $ b_T $, the real-measure distribution $ p(y_T) $ \emph{does} depend on it 
as indicated in Eqs.(\ref{mu_p}). On the other hand,  the value of $ b_T $ is driven, in addition to the current value of
 log-return $ y $ and the model parameters $ \theta $, by the value of the Kramers escape  rate $ \lambda $.  
From a viewpoint of the final expression  (\ref{p_real_measure}), the only role of the machinery developed in the previous sections is to 
compute non-equilibrium corrections $ \sim b_T $ to the real-measure return distribution 
by computing the Kramers escape rate $ \lambda $  in terms of the original model parameters. This can be done by using either Eq.(\ref{Tau_from_Psi}) along with the definition $ \lambda = 1/ \mathcal{T}(y_0) $, or the SUSY QM relation (\ref{E_0_pt}), along with the formula $ \lambda = E_{1}^{-} / h^2 $.
In the practical implementation, we compute $ \lambda $ using Eq.(\ref{Tau_from_Psi}).


\subsection{Examples of calibration to SPX options}
\label{sect_calib_to_SPX_options}

We consider three sets of examples of calibration to market quotes on European options on SPX (the S\&P 500 index). 
In each set, we separately calibrate the model to calls and puts.
Note here that a common approach in the literature is to construct a single risk-neutral implied distribution by assuming the put-call parity and using both call and put option quotes jointly.
While such a procedure would also be possible with the NES model, we prefer to calibrate to calls and puts separately, so that we are able to compare the resulting distributions implied separately by the calls or the puts. One advantage of this approach is that by \emph{not} enforcing the put-call parity, we leave some room for a possible predictable component in returns, see e.g. \cite{Cremers_2010}. Furthermore, a separate calibration of the model to either call or put options is useful as it allows one to separately quantify views of future returns from different segments of the market.

Calibration for all examples considered below is performed using the regularized weighted quadratic loss function:
\beq
\label{loss_option_calib_real}
\mathcal{L}(\theta)  = \frac{1}{N} \sum_{n}^{N}  w_n  \left( C_{NES}(\theta, K_n, S_0) - C_{market}(K_n, S_0) \right)^2 + \lambda \left( \sum_{k=1}^{3}  \omega_k^{(q)}  \mu_k^{(q)}  - r_f +  q + \frac{{h}^2}{2} \right)^2 
\eeq
Here $ C_{market}(K_n, S_0) $  are the quoted prices of SPX options, 
NES option prices $ C_{NES}(\theta, K_n, S_0) $ are computed using Eq.(\ref{opt_price_res}) for call options and a corresponding relation for put options,  
and $ w_n $ are calibration weights. The second term in Eq.(\ref{loss_option_calib_real}) adds Eq.(\ref{expected_return_constraint_2}) as a convex regularization, with the regularization weight $ \lambda $. 
While the first term in the loss function (\ref{loss_option_calib_real}) is non-convex in the model parameters $ \mu, \sigma_1, \sigma_2, a, h $, by increasing the 
value of the regularization $ \lambda $, the loss (\ref{loss_option_calib_real}) can be made convex.  
We choose the calibration weights $ w_n =   \left| \Delta_{BS}(K_n, \sigma_{n}^{impl}, S_0)
\right|^{-1} $ using BS deltas computed using  implied volatilities $ \sigma_{n}^{impl} $ to improve calibration to deep out-of-the money (OTM) options, but in addition also multiply the weight for ATM strikes by a factor of 2 or 3 to retain an accurate calibration to ATM option quotes.

In all three sets of experiments conducted below, for a given option tenor and given pricing date, we calibrate to market quotes on 10 call options and 10 put options.
The strikes are chosen among available market quotes to cover the the range of option deltas between 0.02 and 0.5, in absolute terms, so that we calibrate 
to both ATM strikes and deep OTM strikes. Optimization of the loss function (\ref{loss_option_calib_real}) is done using the \verb|shgo| algorithm available in the Python scientific computing package scipy.

\begin{figure}[]
\begin{center}
\includegraphics[
width=130mm,
height=55mm]{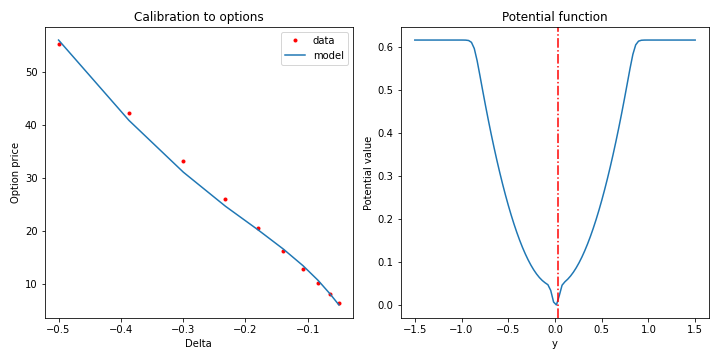}
\caption{Calibration to 1M SPX put options with expiry 08/09/2021 on 07/12/2021and the corresponding implied potential.  The vertical red line corresponds to the current value of log-return $ y_0 $.
} 
\label{calibration_SPX_1M_puts}
\end{center}
\end{figure}

\begin{figure}[]
\begin{center}
\includegraphics[
width=130mm,
height=55mm]{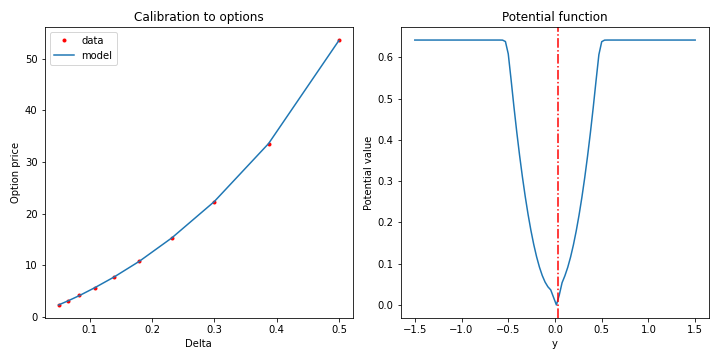}
\caption{Calibration to 1M SPX call options with expiry 08/09/2021 on 07/12/2021 and the corresponding implied potential.  The vertical red line corresponds to the current value of log-return $ y_0 $.
} 
\label{calibration_SPX_1M_calls}
\end{center}
\end{figure}

In the first example, we consider SPX 1M options on 07/12/2021 with maturity on 08/09/2021. The results of calibration are presented in Table \ref{tab_NES_params_1M_SPX} and Figs.~\ref{calibration_SPX_1M_puts} and \ref{calibration_SPX_1M_calls}. Note the difference in implied potentials for puts and calls, as well as different values of inferred model parameters. For both potentials, the current value of the log-return $ y_0 $, as shown by the vertical red lines, 
is located near the bottom of the potential well. This suggests that the price dynamics on this date correspond to an equilibrium regime of small fluctuations around a stable minimum.  

\begin{table}
\begin{center}
  \begin{tabular}{lSSSSSS}
    \toprule
      & {$ \mu $} & {$ \sigma_1 $} & {$ \sigma_2 $} &  {$ a $} & {$ h $}   & { \text{MAPE} } \\
      \midrule
    Puts &0.092 & 0.09 &0.461 &0.505 & 0.159   & 0.035\\
    Calls &0.191 & 0.07 & 0.263 &0.566 & 0.162   & 0.002 \\
    \bottomrule
  \end{tabular}
 \caption{NES parameters obtained by calibration to 10 put and 10 call options on 1M SPX options with expiry 08/09/2021 on 07/12/2021.
 The last column shows the mean absolute pricing errors (MAPE).} 
\label{tab_NES_params_1M_SPX} 
\end{center}  
\end{table}


\begin{figure}[]
\begin{center}
\includegraphics[
width=130mm,
height=55mm]{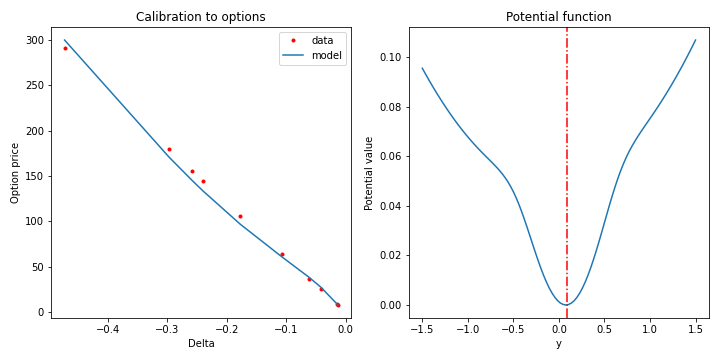}
\caption{Calibration to 1Y SPX put options with expiry 09/21/2021 on 11/06/2020 and the corresponding implied potential. The vertical red line corresponds to the current value of log-return $ y_0 $.
} 
\label{calibration_SPX_1Y_puts}
\end{center}
\end{figure}

\begin{figure}[]
\begin{center}
\includegraphics[
width=130mm,
height=55mm]{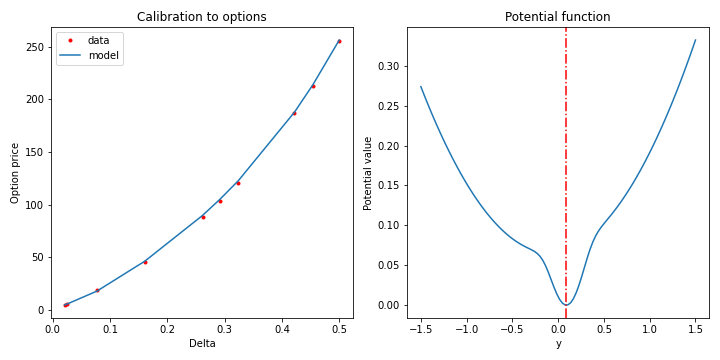}
\caption{Calibration to 1Y SPX call options with expiry  09/21/2021 on 11/06/2020  and the corresponding implied potential.  The vertical red line corresponds to the current value of log-return $ y_0 $.
} 
\label{calibration_SPX_1Y_calls}
\end{center}
\end{figure}

In the second example, we look at longer option tenors, and consider SPX 1Y options on 11/06/2020 with maturity on 09/21/2021. The results of calibration are presented in Table \ref{tab_NES_params_1Y_SPX} and Figs.~\ref{calibration_SPX_1Y_puts} and \ref{calibration_SPX_1Y_calls}. Again, we note the difference in implied potentials for puts and calls, as well as different values of inferred model parameters. Also as in the previous example, for both potentials, the current value of the log-return $ y_0 $ is located near the bottom of the potential well,  suggesting an equilibrium regime of small price fluctuations on this date.  

\begin{table}
\begin{center}
  \begin{tabular}{lSSSSSS}
    \toprule
      & {$ \mu $} & {$ \sigma_1 $} & {$ \sigma_2 $} &  {$ a $} & {$ h $}   & { \text{MAPE} } \\
      \midrule
    Puts &0.092 & 0.251 &0.813 &0.405 & 0.165   & 0.055\\
    Calls &0.106 & 0.123 & 0.505 &0.565 & 0.217   & 0.019 \\
    \bottomrule
  \end{tabular}
 \caption{NES parameters obtained by calibration to 10 put and 10 call options on 1Y SPX options with expiry 09/21/2021 on 11/06/2020. The last column shows the mean absolute pricing errors (MAPE).} 
\label{tab_NES_params_1Y_SPX} 
\end{center}  
\end{table}


\begin{figure}[]
\begin{center}
\includegraphics[
width=130mm,
height=55mm]{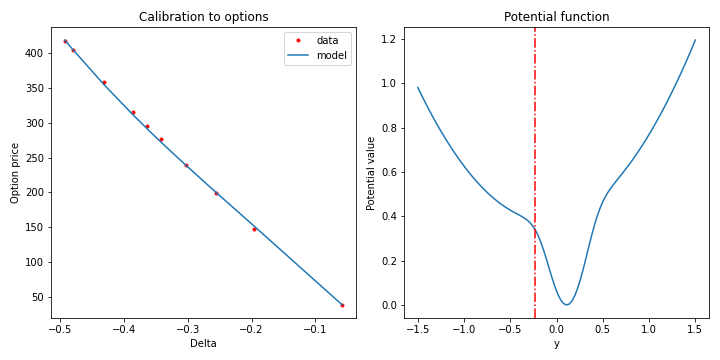}
\caption{Calibration to 6M SPX put options with expiry  09/18/2020 on 03/16/2020 and the corresponding implied potential. The vertical red line corresponds to the current value of log-return $ y_0 $.
} 
\label{calibration_SPX_6M_puts}
\end{center}
\end{figure}

\begin{figure}[]
\begin{center}
\includegraphics[
width=130mm,
height=55mm]{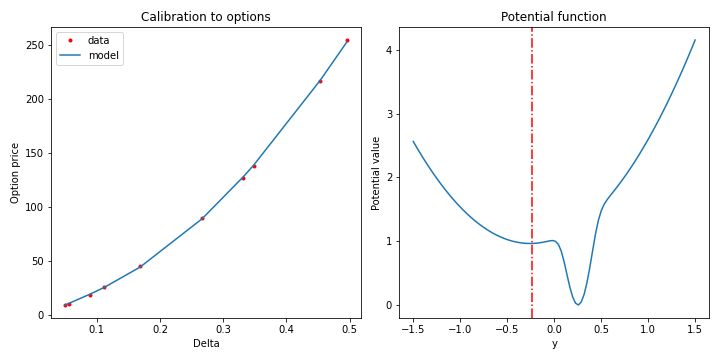}
\caption{Calibration to 6M SPX call options with expiry  09/18/2020 on 03/16/2020 and the corresponding implied potential.  The vertical red line corresponds to the current value of log-return $ y_0 $.
} 
\label{calibration_SPX_6M_calls}
\end{center}
\end{figure}

Finally, we consider an example of a severely distressed market. To this end, we analyze option quotes on 03/16/2020, at the peak of Covid-19 crisis where the SPX index had the largest drop. We consider 6M options with the expiry on 09/18/2020. The results of calibration are presented in Table \ref{tab_NES_params_6M_SPX} and Figs.~\ref{calibration_SPX_6M_puts} and \ref{calibration_SPX_6M_calls}. As in the previous examples, we again note the difference in implied potentials for puts and calls, as well as different values of inferred model parameters. 

\begin{table}
\begin{center}
  \begin{tabular}{lSSSSSS}
    \toprule
      & {$ \mu $} & {$ \sigma_1 $} & {$ \sigma_2 $} &  {$ a $} & {$ h $}   & { \text{MAPE} } \\
      \midrule
    Puts &0.225 & 0.213 &1.120 &0.763 & 0.632   & 0.01\\
    Calls &0.503 & 0.118 & 0.803 &0.662 & 0.824  & 0.007 \\
    \bottomrule
  \end{tabular}
 \caption{NES parameters obtained by calibration to 10 put and 10 call options on 6M  SPX options with expiry  09/18/2020 on 03/16/2020. The last column shows the mean absolute pricing errors (MAPE).} 
\label{tab_NES_params_6M_SPX} 
\end{center}  
\end{table}

\emph{Unlike} the previous examples, for the present case of a distressed market, we note a very different pattern for the current value of the log-return $ y_0 $. 
The current log-return is now located far away from the global minimum of the potential, suggesting strongly non-equilibrium dynamics on that date.  Both sets of put and call market quotes indicate that the value of log-return observed on  03/16/2020 is a strongly 
non-equilibrium initial state, consistently with a prevailing market sentiment on that date.

Interestingly, while both implied potentials suggest that the current value $ y_0 $ is far from the global minimum of the potential and hence describes a non-equilibrium scenario, they differ in the character of a subsequent relaxation mechanism for this initial state. The implied potential for puts in 
 Fig.~\ref{calibration_SPX_6M_puts} 
is a single-well potential, therefore the current state is an unstable state. If the potential itself remains constant through time, this initial state would eventually relax into its global minimum. This would happen even in the limit of zero volatility (zero noise). When the noise is present with $ h > 0 $, it produces both small fluctuations around this minimum, and uncertainty regarding the time needed to reach this minimum.

Differently from the puts, the call options seem to suggest a different type of relaxation dynamics. This is because unlike a single-well potential observed for the puts,  the implied potential for the calls in   Fig.~\ref{calibration_SPX_6M_calls} is a double well potential quite similar to the one shown in the left column in Fig.~\ref{fig_IQED_WF_and_potentials}.
The initial location is in the vicinity of the local minimum corresponding to the left well, while the right well corresponds to the global minimum. As they are separated by the barrrier, the potential in Fig.~\ref{calibration_SPX_6M_calls} describes a scenario of metastability. The relaxation to the global minimum (the true ground state) proceeds via instanton transitions as described in Sect.~\ref{Instantons_and_Kramers}. This suggests that double well potentials implying metastable dynamics may occur when markets are in distress or  
during periods of a high market uncertainty, e.g. due to general elections. In particular, \cite{Gemmill} found a bimodal implied distribution during British elections of 1987 (though not through other British elections in 1992), and suggested that option prices can be used to monitor the market sentiment during elections.

\subsection{Discussion}
\label{sect_discussion}

Gaussian mixtures are frequently used for modeling non-Gaussian market returns on phenomenological grounds, with an objective to fit higher moments of returns such as skewness and kurtosis, see e.g. \cite{Roncalli_2019}.  
This paper shows that with a more theoretically motivated approach, a Gaussian mixture model for returns can instead be imposed as a model of a stationary distribution, which also fixes a non-linear potential in the Langevin dynamics. Using practical simplifications of the NES model presented in this section,  the initial two-component Gaussian mixture that defines a 
square root 
$ \Psi_0 $ of the stationary distribution in Eq.(\ref{Psi_0}) is transformed into non-stationary three-component Gaussian mixtures (\ref{p_real_measure}) and (\ref{GM_3_comp_q}) for the real-world and risk-neutral measures, respectively.   
 The stationary component is parameterized in terms of parameters $ \mu, \sigma_1, \sigma_2, a $. A time-varying, pre-asymptotic component of a fixed-horizon distribution of market returns, is  additionally driven by the volatility parameter $ h $. 

Experiments show that the model is flexible and can 
provide accurate calibration to prices of put and call options on the SPX index for either a benign or stressed market regime. 
While for the former the implied potentials are typically of a single-well type, for a stressed market market environment we observe either a unstable or metastable 
initial state, as in Figs.~\ref{calibration_SPX_6M_puts} and \ref{calibration_SPX_6M_calls}.  
Due to flexibility of the Gaussian mixture model as a model of the ground state $ \Psi_0 $m 
the NES model is able to smoothly shift between stable dynamics with a single minimum, and metastable dynamics obtained when the potential has two local minima separated by a barrier. As illustrated in Fig.~\ref{fig_higher_moments}, the model can  
adapt to different values of skewness and kurtosis  of market returns.

A particularly interesting observation is that when analyzed separately and not jointly as is done in most of related research in the literature, market quotes on the call and put options seem to produce generally different distributions and hence tell different stories about future market dynamics.  
As was mentioned above, this may be explained by a market segmentation between option buyers for calls and puts, and their reliance on possibly different quantitative or internal (mental) models of the world. Because we do not assume put-call parity and do not enforce a joint calibration to puts and calls, 
our framework is less restrictive as it does \emph{not} assume that returns are not predictable \cite{Cremers_2010}. 
Clearly, if the model is to be used to actually predict future returns from option data, it may benefit from \emph{not} assuming that returns are not predictable in the first place.
For such a task, an interesting practical question would be what type of 
option data (calls or puts) is better to use. We leave this question for a future work.

\section{Summary and outlook}
\label{sect_Summary}

McCauley in his insightful and provocative book \cite{McCauley}  pointed out various problems with traditional equilibrium approaches of classical financial models such as e.g. the CAPM model \cite{CAPM} or the Black-Scholes model \cite{BS}. In particular, he argued that fat tails in financial data, which is a matter of everyday concerns of practitioners and academics alike, may be simply the result of applying an unjustified binning process to non-stationary data.
McCauley argued against a neoclassical economic doctrine based on the concept of a market-clearing equilibrium, and showed that among about five different definitions of a market equilibrium commonly used in the mathematical financial modeling literature, none makes sense from the physics' perspective (see also 
\cite{Inverted_World} for related arguments). He also issued a challenge to econophysics as an emerging rival of the classical finance, where research often explicitly or implicitly assumes an equilibrium, potentially suffering from the same problem of a potential data distortion leading to the above-mentioned problem with problematically measured fat tails. 

The Non-Equilibrium Skew (NES) model developed in this paper, as suggested by its name, provides a principled non-equilibrium view of market dynamics.
It follows well defined definitions of equilibrium versus non-equilibrium dynamics as commonly accepted in the mathematical and physical literature.
Equilibrium Langevin dynamics are obtained when a corresponding Langevin potential $ V(y) $ has a single stable minimum. Metastable or unstable dynamics are obtained if the potential $ V(y) $ has multiple minima. Using a two-component Gaussian mixture as the model of a ground state of a quantum mechanical system with a double well potential, we applied methods developed in supersymmetric quantum mechanics (SUSY QM) to find an analytical solution for non-equilibrium, pre-stationary dynamics. 

As non-linear models rarely admit analytical solutions, they usually have much higher computational costs, which is often viewed as a main obstacle preventing their wider adaptation by practitioners. Therefore, availability of an analytical solution for non-linear non-equilibrium market dynamics may be considered a highly desirable property of a model.

Even though the first, SUSY QM-based formulation of the model \emph{is} already analytically tractable, upon further simplifications, we produced a more practical version of the NES model based on insights derived from the previous analysis. The analytical approach  further reduces the computational complexity to a very affordable mixture of three Gaussian distributions, where all parameters are fixed in terms of the original five parameters of the model. The amount of non-stationarity is directly encoded in these Gaussian mixture parameters, and thus can be directly estimated from the model fit to available data, assuming that a data-generating distribution is a non-equilibrium distribution. 
Note that while we only constructed the leading pre-asymptotic corrections to a stationary distribution, higher-order corrections can be computed along the same lines.

To develop a forward-looking way of model estimation, we transformed our resulting Gaussian mixture return distribution into a pricing (risk-neutral) measure, and derived closed-form expressions for prices of index (or stock) options within the NES model. The resulting formulae are given by three-component mixtures of the classical Black-Scholes expressions with different parameters. Again, all parameters in the resulting NES option pricing formula are fixed in terms of the original five model parameters $ \mu, \sigma_1, \sigma_2, a $ and $ h $. 
Computational complexity of the NES model for applications to option pricing is therefore nearly on par with (three times more than) the complexity of the Black-Scholes model - however with all benefits of keeping a principled non-equilibrium approach. With this approach, the question whether market dynamics observed in different regimes can be considered as approximately equilibrium can be answered in a quantitative way, as was illustrated above by analyzing the value of the current log-return value relative to the minimum of the implied potential. 

The NES model demonstrates that a \emph{single volatility parameter} is sufficient to accurately match available market option prices. 
 These results stand in stark contrast to the most of other option pricing models such as local, stochastic, or rough volatility models that need more complex specifications of noise to fit the market data. The approach of this paper offers a viable alternative mechanism based on the analysis of market flows and their impact. As shown in \cite{HD_QED} and discussed in this paper, the interplay of these effects creates a non-linear drift potential in the stock price dynamics. The NES model presented in this paper provides a parameterized model of these dynamics, with drift parameters $  \mu, \sigma_1, \sigma_2, a $ that 
 define the drift potential, and a single volatility parameter $ h $. By fitting these parameters to the market data, we produce implied potentials which replace 
 implied volatility smiles as a way to fit market data. 
          
The NES model can be used for several applications or practical interest. 
In particular, one possible use of this model would be to extract option-implied moments of future returns in order to use them as predictors for actual future returns, volatilities and skewness, and use these signals for portfolio trading. While risk-neutral moments implied by option prices are often analyzed in the literature as predictors of future returns, the NES model lets us extract both risk-neutral and real-measure moments, and thus enrich the set of predictors for such tasks. We can also enlarge the set of market data used for model calibration. In particular, when calibrating to options, in addition to market prices of option, we could take into account open interest data. We could also consider a joint calibration of the NES model to the equity and credit markets data, by adding fitting to 
to market spreads on credit indices such as CDX as proxies to probabilities of large market drops. 
One may expect that such joint calibration would provide a more accurate assessment of the likelihood of large market drops than would be obtained from separate calibrations to the equity and credit markets. For calibration of the NES model to single-stock options, one could consider a similar approach that would use single-name credit default swaps (CDS) instead of credit indices. Another interesting direction of research is to develop a portfolio optimization framework with single stock dynamics as specified in this paper. Such applications and extensions are left here for a future research.     


\appendix

\def\thesection{A}	
\setcounter{equation}{0}
\def\theequation{\thesection.\arabic{equation}}

\section*{Appendix A: Instantons in the Langevin dynamics}
\label{sect:Appendix_A}

In this appendix, we provide the explicit form of the instanton of the Langevin equation  (\ref{Langevin_return_1M}) 
 \beq
 \label{Langevin_return_A}
 d y_t = - \frac{ \partial V (y_t)}{\partial y_t } dt + \sigma d W_t
 \eeq
 using a simple cubic potential
 \beq
 \label{cubic_pot_rho_A}
  V (y_t)  = - \theta y_t + \frac{\kappa}{2} y_t^2 + \frac{g}{3} y_t^3 
\eeq
This potential is chosen to demonstrate a general behavior of instanton solutions in non-linear models. The potential (\ref{cubic_pot_rho_A}) is different from our original potential (\ref{pot_explicit}), as the latter has two local minima, while the former has only one metastable minimum. On the other hand,
the potential (\ref{cubic_pot_rho_A}) captures the presence of a barrier for a particle located near a local minimum of a potential. For potentials with metastability 
such as (\ref{cubic_pot_rho_A}), instantons are defined in a similar way to instantons in models with bistable potentials such as  (\ref{pot_explicit}), as 
solutions corresponding to a tunneling to the `other side' of the potential via tunneling to an equipotential point with the same energy, but \emph{without} a subsequent relaxation to a second, globally stable vacuum of the theory. In quantum field theory, such solutions in metastable models without a subsequent relaxation to a stable minimum are sometimes referred to as \emph{bounces} \cite{Coleman_book}. Though instantons in our main model with the potential  (\ref{pot_explicit}) do not have an explicit solution of terms of elementary functions, they 
behave in a similar way to an instanton solution for the model  (\ref{cubic_pot_rho_A}) to be presented below.  The only difference between them is that the end point of an instanton trajectory for the former is a global minimum of the bistable potential (\ref{pot_explicit}), while for the latter it is an equipotential point of the local minimum in the metastable model (\ref{cubic_pot_rho_A}). 
 
We assume that all parameters are time-independent.
As discussed in \cite{HD_QED}, instantons are solutions satisfying the classical limit of the Langevin equation, but with an \emph{inverted} sign of the potential $ V(\rho) $:
\beq
\label{instanton_eq_ret}
\frac{d y_t}{d t} = \frac{ d V(y_t)}{ d y_t} =  - \theta  + \kappa y_t + g  y_t^2 
\eeq
With the cubic potential (\ref{cubic_pot_rho_A}), its derivative appearing in the right hand side of Eq.(\ref{instanton_eq_ret}) is a second order polynomial.
It has two roots, one corresponding to a local minimum, and another corresponding to a maximum. We denote the local minimum and maximum positions as $ y_{\star} $ and $ y^{\star} $, respectively. 
A \emph{reflection point} $ \hat{y} $ is a position on the other side of the barrier that matches the value $ V(y_{\star}) $ at the local minimum, so that the values 
 $ y_{\star} $ and $ \hat{y} $ are equipotential points. In classical mechanics, a motion between two equipotential points can proceed with vanishing kinetic energy, i.e. infinitely slow.   

The instanton solution for the potential (\ref{cubic_pot_rho_A}) can be found by directly integrating 
Eq.(\ref{instanton_eq_ret}):
\beq
\label{QED_instanton_ret_1D}
y_t = y_{\star} + \frac{ \hat{y}- y_{\star}}{1 +  e^{ - g(y_{\star} - \hat{y}) (t - t_c)}}
\eeq
where parameter $ t_c $ is called the instanton center.
This is a shifted and re-scaled logistic law (sigmoid function) in time. The value $ y_0 $ of the instanton solution at time $ t = 0 $ is 
determined by the position $ t_c $ of the instanton center.\footnote{This is different from conventional initial-value problems where an initial condition at time $ t = 0 $ is an input to the problem. For the instanton, there are only fixed initial and terminal values $ y_{\star} $ and  $ \hat{y} $ for $ t = \pm \infty $, while the value at $ t = 0 $ is determined by the instanton center $ t_c $ rather than being an independent input.} In particular, if $ t_c = 0 $, then  $ y_0  = (y_{\star} +  \hat{y})/2 $. 
Assuming that $y_{\star}  >  \hat{y} $, the solution at $ t \rightarrow - \infty $ is  $ \rho_{-\infty} = \rho_{\star} $, and as $ t \rightarrow  \infty $,  it approaches $ y_{\infty} =  \hat{y} $. 
The instanton therefore starts at the minimum $ y_{\star} $ at $ t  \rightarrow  - \infty $, and then climbs the potential wall all the way to the maximum 
$ y^{\star} $ , going \emph{along} the gradient of the potential, and not against it as it would only be allowed in classical mechanics. After reaching the maximum 
$ y^{\star} $ at some time, the instanton then moves to the reflection point $ \hat{y} $, reaching it at time $ t \rightarrow \infty $. On this part of its trajectory, the instanton moves \emph{against} the gradient, consistently with conventional laws of mechanics.
We can also consider an \emph{anti-instanton} that lives backward in time: it starts at the reflection point  $ \hat{y}$ at  $ t  \rightarrow  - \infty $ and reaches the local minimum
at time $ t  \rightarrow  \infty $. 

Note that while $ y_{\star} $ and $ \hat{y} $ give the limiting positions at $ t \rightarrow - \infty $ and $ t = \infty $, the transition between them is well localized in time around the instanton center $ t_c $. 

\def\thesection{B}
\setcounter{equation}{0}
\def\theequation{\thesection\arabic{equation}}

\section*{Appendix B: SUSY}
\label{sect_Appendix_SUSY}

The Hamiltonian $ \mathcal{H}_{-} $ transforms into 
the partner Hamiltonian $ \mathcal{H}_{+} $ if we flip the sign of 
the potential $ V(y) \rightarrow - V(y) $. They can be paired in the following
matrix-valued Hamiltonian:
\bea
\label{matrix_H}
\mathcal{H} = 
 \left[ \begin{array}{cc}
  \mathcal{H}_{+} & 0   \\
  0 & \mathcal{H}_{-} \\
\end{array} \right]
= 
 \left[ \begin{array}{cc}
  \mathcal{A} \mathcal{A}^{+}  & 0   \\
  0 & \mathcal{A}^{+} \mathcal{A} \\
\end{array} \right].
\eea
This is the Hamiltonian of the Euclidean supersymmetric quantum mechanics of Witten
\cite{Witten_SUSY}. It can also be represented in a form that involves fermion (anti-commuting) fields $ \psi_t, \psi_t^{+} $, in addition to the conventional boson (i.e., commuting) field 
$ y_t $, see e.g. \cite{Junker}. Alternatively, two purely boson Hamiltonians 
$ \mathcal{H}_{-} = \mathcal{A}^{+} \mathcal{A} $ and 
$ \mathcal{H}_{+} = \mathcal{A} \mathcal{A}^{+} $ can be thought of as representing two different fermion sectors of the model. The potential
$ V(y) $ is referred to in the context of SUSY models as the superpotential. For a brief review of SUSY quantum mechanics, see e.g. \cite{Junker, SUSY_QM_Cooper}.

Instead of representation in terms of operators $ \mathcal{A}, \, \mathcal{A}^{+} $,
we can equivalently express the Hamiltonian (\ref{matrix_H}) in terms of supercharges 
\bea
\mathcal{Q}_1 = 
\frac{1}{\sqrt{2}}\left[ \begin{array}{cc}
    0 & \mathcal{A}    \\
   \mathcal{A}^{+} & 0  \\
\end{array} \right], \; \; \; 
\mathcal{Q}_2 = 
\frac{i}{\sqrt{2}}\left[ \begin{array}{cc}
    0 & -\mathcal{A}    \\
   \mathcal{A}^{+} & 0  \\
\end{array} \right]
\eea
This gives
\beq
\label{H_Q12}
\mathcal{H} = 2 \mathcal{Q}_1^2 = 2 \mathcal{Q}_2^2 
= \mathcal{Q}_1^2 +  \mathcal{Q}_2^2.
\eeq
This means that the Hamiltonian $ \mathcal{H} $ commutes with both 
supercharges, i.e. $ [\mathcal{H},\mathcal{Q}_1] \equiv \mathcal{H} \mathcal{Q}_1
- \mathcal{Q}_1 \mathcal{H}  = 0 $, and 
$ [\mathcal{H},\mathcal{Q}_2] = 0 $, therefore 
the supercharges $ \mathcal{Q}_1, \mathcal{Q}_2 $ are constants in time. Furthermore,
Eq.(\ref{H_Q12}) shows that eigenvalues of both Hamiltonians 
$ \mathcal{H}_{\pm} $ are non-negative, with zero being the lowest 
possible eigenvalue.

Due to the factorization property (\ref{factorization}), if $ \Psi_n^{-} $ is an eigenvector of $ \mathcal{H}_{-} $ with an eigenvalue $ E_n^{-} > 0 $ (where
$ n = 1, 2, \ldots $), than the state $ \Psi_n^{+} \equiv \left(E_n^{-} \right)^{-1/2} \mathcal{A}  \Psi_n^{-} $ will be an eigenstate of the SUSY partner Hamiltonian  $ \mathcal{H}_{+} $ with the same eigenvalue (energy) $ E_n^{-} $ (the factor $ \left(E_n^{-} \right)^{-1/2} $ is introduced here for a correct normalization.) This is seen from the following transformation
\beq
\label{SUSY_H}
\mathcal{H}_{+}  \Psi_n^{+} 
= \left(E_n^{-} \right)^{-1/2} \mathcal{A} \mathcal{A}^{+} \mathcal{A}   \Psi_n^{-}  
= \left(E_n^{-} \right)^{-1/2} \mathcal{A}  \mathcal{H}_{-} \Psi_n^{-} 
= \left(E_n^{-} \right)^{-1/2} \mathcal{A} E_n^{-} \Psi_n^{-} 
= E_n^{-} \Psi_n^{+}, \; \; n = 1, 2, \ldots,  \nonumber
\eeq
which means that all eigenstates of  spectra of $ H $, except 
possibly for a 'vacuum' state with energy $ E_0^{-} = 0 $, should be degenerate in energy with eigenstates of the SUSY partner Hamiltonian  $ \mathcal{H}_{+} $ \cite{Witten_SUSY}. Such zero-energy ground state would be unpaired, while all higher states would be doubly degenerate between the SUSY partner Hamiltonians 
$ \mathcal{H}_{\pm} $:
\bea
\label{SUSY_relations}
&&  \mathcal{H}_{-} \Psi_0^{-} = \mathcal{A}  \Psi_0^{-} = 0, \; \;   E_0^{-} = 0 \nonumber \\
&& \Psi_{n+1}^{-} = \left(E_n^{+} \right)^{-1/2}
\mathcal{A}^{+} \Psi_{n}^{+}, \; \; \; 
\Psi_{n}^{+} = \left(E_{n+1}^{-} \right)^{-1/2}
\mathcal{A} \Psi_{n+1}^{-}, \; \; \; n = 0, 1, \ldots \\
&& E_{n+1}^{-} = E_{n}^{+}, \; \; \; n = 0, 1, \ldots 
\nonumber 
\eea

The existence or non-existence of a zero-energy ground state
$ E_0^{-} = 0 $ has to do with supersymmetry being unbroken or 
spontaneously broken.
In scenarios with spontaneous breaking of SUSY, supersymmetry is a symmetry of a Hamiltonian but not of a ground state of that Hamiltonian. On the other hand, an unbroken SUSY is characterized by the existence of a normalizable ground state $ \Psi_0 $ with strictly zero energy $ E_0 = 0 $, while for a spontaneously broken SUSY the energy of the ground state is larger than zero \cite{Witten_SUSY}:
\bea
\label{Unbroken_SUSY}
\mbox{Unbroken SUSY:}  &&  A  \Psi_0 = 0  \cdot \Psi_0  = 0 \; \; (E_0 = 0) \nonumber \\
\mbox{Spontaneously broken SUSY:}  &&  A  \Psi_0  = E_0   \Psi_0 , \; \; E_0 > 0. 
\eea
For SUSY to be unbroken, the derivative of the superpotential 
$ V'(y) = \partial V/ \partial y $ should have different signs at $ y = \pm \infty $, which means that that it should have an odd number of zeros at real values 
of $ y $. 


\def\thesection{C}	
\setcounter{equation}{0}
\def\theequation{\thesection.\arabic{equation}}

\section*{Appendix C: Logarithmic Perturbation Theory}
\label{sect_Appendix_LPT}

This appendix presents Logarithmic Perturbation Theory (LPT)
for a ground state wave function and energy for 
a one-dimensional Schr{\"o}dinger equation. Further details can be found 
in \cite{LPT, Turbiner}.

We consider the following Schr{\"o}dinger equation with $ m = 1 $:
\beq
\label{pertproblem}
\left[ - \frac{\hbar^2}{2} \frac{\partial^2}{\partial x^2}  + V_0 (x) + \alpha V_{1} (x)
\right] \psi (x) = E \psi(x)
\eeq
where parameter $ \alpha $ is assumed to be small, and we are only interested in the lowest energy eigenstate. 
As the ground 
state function is nodeless, we look for a solution in the form
\beq
\label{ansatz_LPT}
\psi(x) = \exp \left[ - \frac{G(x)}{\hbar} \right]
\eeq
where a function $ G(x) $ is assumed to be twice differentiable. Substituting
(\ref{ansatz_LPT}) into (\ref{pertproblem}), we obtain the Riccati equation for
the derivative $ g(x) = G'(x) $:
\beq
\label{Riccati}
 \frac{1}{2} g^2(x) - \frac{\hbar}{2} g'(x)  +  E - V_0 (x) - \alpha V_1 (x) = 0
\eeq
We can solve this equation in terms of 
perturbative series for $ g(x) $ and $ E $:
\beq
\label{pertseries}
g(x) = \sum_{k =0}^{\infty} \alpha^{k} g_k (x) \, , \; \; \; 
E =  \sum_{k=0}^{\infty} \alpha^{k} \bar{E}_k
\eeq
Note that here $ k $ enumerates the perturbative order of the calculation, and not the energy level. As we are only interested here in the lowest energy eigenstate,
we do not need a subscript labeling different eigenstates. Using 
(\ref{pertseries}) in (\ref{Riccati}) and comparing coefficients for various 
powers of $ \lambda $, we find
\bea
\label{LPTbasic}
 \frac{1}{2} g_{0}^{2} - \frac{\hbar}{2}  g_{0}' &=&  V_0  - \bar{E}_0  
\nonumber \\
g_{0} g_{1} - \frac{\hbar}{2}  g_{1}'  &=&  V_1  - \bar{E}_1 \\
2 g_{0} g_{k} - \hbar g_{k}' &=&   
\sum_{j = 1}^{k-1} g_j g_{k-j} - \bar{E}_k \;, \; \; \; k \geq 2 \nonumber
\eea
The first equation here is just the Schr{\"o}dinger equation for the 
unperturbed problem with the ground-state WF $ \psi_0(x) $, expressed in terms of $ g_0(x) $.  Multiplying the second equation by the integrating factor
$ \exp(-2 G_0(x)/ \hbar ) = \psi_0^2(x)$, we obtain
\beq
\label{integratingfactor}
\frac{d}{dx} \left( g_1 (x) \psi_{0}^{2}(x) \right) = - \frac{2}{\hbar} \left(V_1 (x)  - \bar{E}_1 \right) \psi_{0}^{2}(x)
\eeq
Integrating this equation from $ -\infty $ to $ \infty $ and using the fact that 
$ \psi_{0}(x) $ satisfies the zero boundary
conditions at $ x = \pm \infty $, we find the first order correction 
to the energy:
\beq
\label{energy1order}
\bar{E}_1 = \frac{\int_{-\infty}^{\infty} dz \, V_1 (z) \psi_{0}^{2}(z) }{
\int_{-\infty}^{\infty} dz  \, \psi_{0}^{2}(z) }
\eeq
Now when $ E_1 $ is specified,  Eq.(\ref{integratingfactor}) can be integrated from $ - \infty $ to $ x $ for $ x > 0 $, or from $ x $ to $ \infty $ for $ x < 0 $, producing the following result:
\beq
\label{dWF1order}
g_1 (x) = 
\left\{ \begin{array}{cc} 
 \frac{2}{ \hbar}  \frac{1}{\psi_{0}^{2} (x) } \int_{-\infty}^{x} dz \, 
\left( \bar{E}_1  - V_1 (z) \right) \psi_{0}^{2}(z), \; \; \; &  x > 0   \\
-  \frac{2}{ \hbar}  \frac{1}{\psi_{0}^{2} (x) } \int_{x}^{\infty} dz \, 
\left( \bar{E}_1  - V_1 (z) \right) \psi_{0}^{2}(z),  \; \; \;   &  x < 0  \\
\end{array} \right.   
\eeq
Using this expression, we obtain the first order correction to the 
wave function:
\beq
\label{WF1order}
G_1 (x) = 
\left\{ \begin{array}{cc} 
\frac{2}{ \hbar}  \int_{-\infty}^{x} dy  \frac{1}{\psi_{0}^{2} (y) } \int_{-\infty}^{y} dz \, 
\left( \bar{E}_1  - V_1 (z)\right) \psi_{0}^{2}(z) + C, \; \; \; &  x > 0   \\
- \frac{2}{ \hbar}   \int_{-\infty}^{x} dy  \frac{1}{\psi_{0}^{2} (y) } \int_{y}^{\infty} dz \, 
\left( \bar{E}_1  - V_1 (z) \right) \psi_{0}^{2}(z) + C,  \; \; \;   &  x < 0  \\
\end{array} \right. 
\eeq
Here $ C  $ is a constant that should be fixed from 
a normalization condition on a perturbed wave function. 
Note that the integrand in this expression is non-singular at 
$ y \rightarrow \infty $ due to Eq.(\ref{energy1order}).

We can further introduce the ``effective potential'' which is calculated 
recursively at each given order in $ \lambda $:
\beq
\label{effpot}
V_{i}^{eff} (x) = \sum_{j = 1}^{i-1} g_j (x) g_{i-j} (x), \; \; \; i \geq 2
\eeq
Performing the same manipulations as above 
with the last of Eqs.(\ref{LPTbasic}), we obtain
\bea
\label{anyorder}
\bar{E}_i &=& \frac{\int_{-1}^{1} dz \, 
 V_{i}^{eff} (z) \psi_{0}^{2}(z)}{
\int_{-1}^{1} dz  \psi_{0}^{2}(z) }   \nonumber \\
g_i (x) &=& \frac{1}{ \hbar \psi_{0}^{2} (x) } \int_{-1}^{x} dz \, 
\left[ \bar{E}_i -  V_{i}^{eff} (z) \right] \psi_{0}^{2}(z) \\
G_i (x) &=& \frac{1}{\hbar} \int_{-1}^{x} dy \frac{1}{\psi_{0}^{2}(y)}
\int_{-1}^{y} dz \left[ \bar{E}_i - V_{i}^{eff} (z) \right] \psi_{0}^{2}(z) + C_i 
\eea
Equations (\ref{energy1order})-(\ref{anyorder}) provide a 
recursive scheme by which perturbative corrections to the energies and 
wave functions can be calculated to any order in the coupling $ \lambda $.
In contrast to the standard Rayleigh-Schr{\"o}dinger (RS) perturbation theory, 
they do not require a 
calculation of matrix elements of the perturbation potential $ V_1 $. While
the new scheme can be shown to be equivalent to the RS perturbation 
theory \cite{LPT}, it is far more convenient for calculating perturbative 
corrections due to its quadrature, recursive form.

\def\thesection{D}
\setcounter{equation}{0}
\def\theequation{\thesection\arabic{equation}}

\section*{Appendix D: Higher moments of Gaussian mixtures}

\label{sect_Appendix_D}

Consider a random real-valued variable $ X $ whose probability density function (pdf) of realizations $ X =  x $ is given by a Gaussian mixture
\beq
\label{Gaussian_mixture_model}
f_{\boldsymbol{\theta}}(x) = \sum_{k=1}^{K} \omega_k \phi( x | \mu_k, \sigma_k^2), \; \; \;   \sum_{k=1}^{K} \omega_k = 1, \; \; \; 0 \leq  \omega_k \leq 1
\eeq
Here $ \boldsymbol{\theta} $ stands the vector of all parameters $ ( \mu_k, \sigma_k^2) $ of  $ K $ Gaussian components, with 
$ \phi( x |\mu_k, \sigma_k^2) $ being the Gaussian pdfs with the means $ \mu_k $ and variances $  \sigma_k^2 $.
The moment generating function (MGF) is defined as follows:
\beq
\label{MGF}
M(z) := \mathbb{E}\left[e^{  z X } \right] = \sum_{k=1}^{K}  \omega_k   \mathbb{E}_k \left[e^{  z  X } \right]  = 
 \sum_{k=1}^{K}  \omega_k  \exp\left[  \mu_k z + \frac{\sigma_k^2}{2} z^2 \right]
\eeq
where $ \mathbb{E}_k \left[ \cdot \right]  $ stands for an expected value computed using the $ k $-th component of the Gaussian mixture (\ref{Gaussian_mixture_model}). All moments $ M_n $ can be conveniently computed from the MGF (\ref{MGF}) as coefficients of its Taylor expansion around $ z = 0 $:
\beq
\label{MGF_Taylor}
M(z) = \sum_{n=0}^{\infty} \frac{M_n}{n!} z^n
\eeq
This gives uncentered moments
\bea
\label{uncentered_moments}
&& \hat{\mu} := M_1 = \mathbb{E} \left[ X \right] = \sum_{k=1}^{K} \omega_k \mu_k, \; \; \; 
M_2 = \mathbb{E} \left[ X^2 \right] = \sum_{k=1}^{K} \omega_k \left(\mu_k^2 + \sigma_k^2 \right),  \\
&& M_3 = \mathbb{E} \left[ X^3 \right] = \sum_{k=1}^{K} \omega_k \left(\mu_k^3 + 3 \mu_k \sigma_k^2 \right), \; \; \; 
M_4 = \mathbb{E} \left[ X^4 \right] = \sum_{k=1}^{K} \omega_k \left(\mu_k^4 + 6 \mu_k^2 \sigma_k^2 + 3  \sigma_k^4 \right) \nonumber
\eea
Centered moments can be computed using (\ref{uncentered_moments}). In particular, for the variance of the mixture we obtain
\beq
\label{var_mixture}
\hat{\sigma}^2 := \mathbb{E} \left[ X^2 \right]  -  \left( \mathbb{E} \left[ X \right] \right)^2 = \sum_{k=1}^{K} \omega_k \left(\mu_k^2 + \sigma_k^2  - 
\mu_k \sum_{k'} \omega_{k'} \mu_{k'} \right) 
\eeq
The skewness is a normalized centered third moment
\beq
\label{skewness}
\tilde{\mu}_3 := \mathbb{E} \left[ \left( \frac{X - \hat{\mu}}{\hat{\sigma}} \right)^3 \right] = 
\frac{ \mathbb{E}\left[\left( X - \hat{\mu} \right)^3 \right]}{ \left( \mathbb{E}\left[ \left( X - \hat{\mu} \right)^2 \right] \right)^{3/2}} 
= \frac{ \mathbb{E}\left[X^3 \right] - 3 \hat{\mu}  \mathbb{E} \left[X^2 \right]  + 2 \hat{\mu}^3}{ \hat{\sigma}^{3}} 
\eeq
The kurtosis is the normalized centered fourth moment
\beq
\label{kurtosis}
\tilde{\mu}_4 := \mathbb{E} \left[ \left( \frac{X - \hat{\mu}}{\hat{\sigma}} \right)^4 \right] = 
\frac{ \mathbb{E}\left[\left( X - \hat{\mu} \right)^4 \right]}{ \left( \mathbb{E}\left[ \left( X - \hat{\mu} \right)^2 \right] \right)^{2}} 
= \frac{ \mathbb{E}\left[X^4 \right] - 4 \hat{\mu}  \mathbb{E}\left[X^3 \right] + 6 \hat{\mu}^2   \mathbb{E}\left[X^2 \right] - 3 \hat{\mu}^4}{ \hat{\sigma}^{4}} 
\eeq


\begin{thebibliography}{99}

\bibitem{LPT} Y. Aharonov and C.K. Au, ``New Approach to Perturbation 
Theory'', {\it Phys. Rev. Lett.} {\bf 42} (1979) 1582. 
\bibitem{Alexander} C.~Alexander and S.~Narayanan, ``Option Pricing with Normal Mixture Returns: Modelling Excess Kurtosis and Uncertainty in Volatility",
 ICMA Centre Discussion Papers in Finance icma-dp2001-10, Henley Business School, Reading University (2001).
 \bibitem{BS} F.~Black and M.~Scholes, "The Pricing of Options and Corporate Liabilities", Journal of Political Economy, 
Vol. 81(3),  637-654, 1973.
\bibitem{BC}  J.P.~Bouchaud and R.~Cont, ``A Langevin Approach To Stock Market", {\it The European Physical Journal B}, {\bf 6}(4), 543-550 (1998). 
\bibitem{Brown} M.~Bernstein and L.S.~Brown, "Supersymmetry and the Bistable Fokker-Planck Equation", {\it Physical Review Letters}, {\bf 52} (22), 1933-1935 
(1984).
\bibitem{Chabi-Yo-2007} F.~Chabi-Yo, D.~Liesen, and E.~Renault, ``Implications of Asymmetry Risk for Portfolio Analysis and Asset Pricing", Bank of Canada (2007).
 \bibitem{Coleman_book} S.~Coleman, {\it Aspects of Symmetry. Selected Erice Lectures}, Cambridge University Press (1988).
\bibitem{SUSY_QM_Cooper} F.~Cooper, A.~Khare, and I.~Sukhatme, {\it Supersymmetry in Quantum Mechanics}, World Scientific (2001).
\bibitem{Corrado}  C.J.~Corrado and  T.~Su, ``Implied volatility skews and stock index skewness and kurtosis implied by S\&P 500 index option prices", Journal of Derivatives 5, 8-19 (1996).
\bibitem{Cremers_2010} M.~Cremers and D.~Weinbaum, ``Deviations from Put-Call Parity and Stock Return Predictability", Journal of Financial and Quantitative 
Analysis, vol. 45, no.2, pp. 335-367 (2010).
\bibitem{SUSY_tunneling_asymmetric} A.~Gangopadhyaya, P.K.~Panigrahi, and U.P.~Sukhatme, ``Supersymmetry and Tunneling in an Asymmetric Double Well", {\it Phys. Rev. A}, vol. 47 (no. 4), 2720-2724 (1993). 
\bibitem{Gardiner} Gardiner, {\it Handbook of Stochastic Methods},  Springer (1996). 
\bibitem{Gemmill} G.~Gemmill and A.~Saflekos, ``How Useful are Implied Distributions? Evidence from Stock-Index Options" (2000), The Journal of Derivatives, 7 (3), 83-91 (2000), available at  https://www.bis.org/publ/bisp06e.pdf. 
\bibitem{Feigelman_Tsvelik} M.V.~Feigelman and A.M.~Tsvelik, "Hidden Supersymmetry of Stochastic Dissipative Dynamics",  Sov. Phys. JEPT, {\bf 56} (4), 823-830 (1982).
\bibitem{Ferreira_2018} T.R.T.~Ferreira, ``Stock Market Cross-Sectional Skewness and
Business Cycle Fluctuations", Board of Governors of the Federal Reserve System, https://www.federalreserve.gov/econres/ifdp/files/ifdp1223.pdf (2018).
\bibitem{Hanggi_1986} P.~Hanggi, ``Escape from a Metastable State", {\it Journal of Statistical Physics}, {\bf 42} (1/2) 105-148 (1986).  
\bibitem{HD_QED} I.~Halperin and M.F.~Dixon, ``Quantum Equilibrium-Disequilibrium:  Asset Price Dynamics, Symmetry Breaking, and Defaults as Dissipative Instantons", {\it Physica A}  {\bf 537}, 122187,  https://doi.org/10.1016/j.physa.2019.122187 (2020).
\bibitem{Inverted_World} I.~Halperin, ``The Inverted World of Classical Quantitative Finance: a Non-Equilibrium and Non-Perturbative Finance Perspective",
https://arxiv.org/abs/2008.03623 (2020).
\bibitem{Harvey_2000} C.R.~Harvey and A.~Sidduque, ``Time-Varying Conditional Skewness and the Market Risk Premium", Research in Banking in Finance, vol. 1, pp. 25-58 (2000).
\bibitem{KKS} W.-Y. Keung, E. Kovacs, and U. Sukhatme, ``Supersymmetry and
Double Well Potentials'', {\it Phys. Rev. Lett.} {\bf 60} (1988) 41.  
\bibitem{Junker} G.~Junker, {\it Supersymmetric Methods in Quantum and Statistical Physics}, Springer 1996.
\bibitem{Bouchaud_Skew_risk_premia} Y. ~Lemp{\' e}ri{\' e}re, C.~Deremble, T. T. ~Nguyen, P. ~Seager, M. ~Potters, J. P. ~Bouchaud, ``
Risk Premia: Asymmetric Tail Risks and Excess Returns", Quantitative Finance vol. 17, no.1, 1-14, https://arxiv.org/abs/1409.7720 (2017).
\bibitem{Landau_QM} L.D.~Landau and E.M.~Lifschitz, {\it Quantum Mechanics}, Elsevier (1980).
\bibitem{Langevin} P.~Langevin, ``Sur la Th{\'e}orie du Mouvement Brownien", {\it Comps Rendus Acad. Sci.} (Paris) 146, 530-533 (1908).
\bibitem{Roncalli_2019} E.~Lezmi, H.~Malongo, T.~Roncalli, and R.~Sobotka, ``Portfolio Allocation with Skewness Risk: a Practical Guide", available at 
http://www.thierry-roncalli.com/download/Skewness-Risk-Allocation.pdf.
\bibitem{McCauley} McCauley, Dynamics of Markets: The New Financial Economics, 2nd edition, Cambridge University Press (2009).
\bibitem{Ren_2007} C.~Ren and A.R.~MacKenzie, ``Closed-Form Approximations to the Error and Complimentary Error Functions and Their 
Applications in Atmospheric Science", Atmospheric Science Letters, v. 8, issue 3, pp.70-73 (2007),  https://rmets.onlinelibrary.wiley.com/doi/full/10.1002/asl.154.
\bibitem{CAPM} W.F.~Sharpe, ``Capital Asset Prices: A Theory of Market Equilibrium Under Conditions of Risk", {\it Journal of Finance}, {\bf 19} (3), 425?442 (1964).
\bibitem{Stilger} P.S.~Stilger, A.~Kostakis, and S.H.~Poon, ``What Does Risk-Neutral Skewness Tell Us About Future Stock Returns?", Management Science, vol. 63, Issue 3, 1657-2048 (2016).
\bibitem{Sornette_book} D.~Sornette, {\it Why Stock Markets Crash}, Princeton University Press (2003). 
\bibitem{Turbiner} A.V. Turbiner, ``A New Approach to the Eigenvalue Problem in
Quantum Mechanics: Convergent Perturbation Theory for Rising Potentials'', 
{\it JETP Lett.} {\bf 30} (1979) 352, {\it J.Phys.} {\bf A14} (1981)1641. 
\bibitem{vanKampen} N.G.~Van Kampen, {\it Stochastic Processes in Physics and Chemistry}, North-Holland (1981).
\bibitem{Witten_SUSY} E.~Witten, "Dynamical Breaking of Supersymmetry", {\it Nuclear Physics B} {\bf 188}(3-5), 513-554 (1981).
\bibitem{Zinn-Justin-QFT} C.~Zinn-Justin, {\it Quantum Field Theory and Critical Phenomena}, Fourth Edition, Clarendon Press, Oxford (2002).


\end{thebibliography}
\end{document}